\documentclass{article}
\def\version{29.3.2013 -{\color{blue}\tiny NO DAMAGE}}
\def\users{world}
\usepackage{amsmath,color}

\date{}
\newtheorem{theorem}{Theorem}[section]

\newtheorem{lemma}[theorem]{Lemma}
\newtheorem{proposition}{Proposition}

\newtheorem{definition}[theorem]{Definition}
\newtheorem{remark}{Remark}
\title{Thermomechanics of hydrogen storage in metallic hydrides: modeling and analysis}

\author{Toma\v{s} Roub\'i\v{c}ek and Giuseppe Tomassetti}

 \def\emaila{Email:\tt tomas.roubicek@mff.cuni.cz}
 \def\emailb{Email:\tt tomassetti@ing.uniroma2.it}

\def\thanks{Fruitful discussions with {}Elena Bonetti, 
{}Michal Pavelka, {}Paolo Podio-Guidugli, 
{}and Ulisse Stefanelli{} as well as the 
comments of two anonymous referees {}are warmly acknowledged.
This work was 
supported by grants 201/09/0917,
201/10/0357, and 105/13/18652S (GA \v CR) 
and
the CENTEM project no.\ CZ.1.05/21.00/03.0088 
(cofounded from ERDF within the OP RDI programme, M\v SMT \v CR) at 
New Technologies Research Centre (Univ.\ of West Bohemia, Plze\v n).}

\newcommand\mathbbmss{\mathsf}
\newcommand\mathbbm{\mathbb}

\usepackage{amsfonts}
\usepackage{amssymb}
\usepackage{mathrsfs} %needed for script fonts
\usepackage[normalem]{ulem} %for insert-replace macros. the normalem option keeps italics

\newcommand\kt{^k_\tau}

\newcommand{\e}{\varepsilon}
\renewcommand{\r}{\varrho}

\renewcommand{\d}{{\rm d}}

\newcommand{\R}{\mathbb R}

\newcommand{\N}{\mathbb N}

\newcommand{\bbC}{\mathbb C}

\newcommand{\bbD}{\mathbb D}

\renewcommand{\O}{{\Omega}}

\newcommand{\QED}{\ \hfill$\ \hfill\Box$}

\definecolor{gray}{rgb}{0.9,0.9,0.9}
\newcommand\DT[1]{\mathchoice
                 {{\buildrel{\hspace*{.1em}\text{\LARGE.}}\over{#1}}}
                 {{\buildrel{\hspace*{.1em}\text{\Large.}}\over{#1}}}
                 {{\buildrel{\hspace*{.1em}\text{\large.}}\over{#1}}}
                 {{\buildrel{\hspace*{.1em}\text{\large.}}\over{#1}}}}
\newcommand\DDT[1]{\mathchoice
   {{\buildrel{\hspace*{.1em}\text{\LARGE.\hspace*{-.1em}.}}\over{#1}}}
   {{\buildrel{\hspace*{.1em}\text{\Large.\hspace*{-.1em}.}}\over{#1}}}
   {{\buildrel{\hspace*{.1em}\text{\large.\hspace*{-.1em}.}}\over{#1}}}
   {{\buildrel{\hspace*{.1em}\text{\large.\hspace*{-.1em}.}}\over{#1}}}}

\newcommand\dx{\ \!{\rm d}x}
\newcommand\dS{\ \!{\rm d}S}

\newcommand\bfalpha{{\boldsymbol\alpha}}
\newcommand\bfeps{{\boldsymbol\varepsilon}}

\newcommand\bfz{\mathbf z}

\renewcommand\kt{^{k}_\tau}

\newcommand\kkt{^{k-1}_\tau}

\newcommand\bbv{\mathbbmss v}
\newcommand\bbs{\mathbbmss s}
\newcommand\bbd{\mathbbmss d}

\newcommand\bfsigma{\boldsymbol\sigma}

\newcommand\bfh{\mathbf h}

\newcommand\bbS{\mathbbmss S}
\newcommand\bfn{\mathbf n}
\newcommand\bfq{\mathbf q}

\newcommand\bbc{\chi}
\newcommand\CHI{\mathbbmss c}
\newcommand\bbm{\bbc}
\newcommand\bfepstr{E}
\newcommand\dtt{\mathfrak d}
\usepackage{mathrsfs}
\renewcommand\mathcal{\mathscr}
\newcommand{\bbf}{\mathsf g}
\newcommand\bff{\mathsf f}
\newcommand\bfu{\mathsf u}
\newcommand\bfv{\mathsf v}
\newcommand\bbM{\mathbb M}
\newcommand\bbK{\mathbb K}

\usepackage{ifthen}
\ifthenelse{\equal{\users}{world}}{}
{
\usepackage{fancyhdr}
\pagestyle{fancy}
\headheight=28pt
\definecolor{grey}{rgb}{0.6,0.6,0.6}
\rhead{\color{grey}
Hydrogen storage in hydrides. \\ by T.Roubicek \& G.Tomassetti}
\chead{}
\lhead{\color{grey}Version \version, file: \jobname.tex\\compiled: 
\number\day.\number\month.\number\year\ at
\the\hour:\ifnum\minute<10 0\fi\the\minute\ h }
}
\newcount\hour \newcount\minute
\hour=\time 
\divide \hour by 60
\minute=\time
\loop \ifnum \minute > 59 \advance \minute by -60 \repeat

\usepackage{ulem}
\usepackage{cancel}
\ifthenelse{\equal{\users}{world}}{

	\newcommand{\DELETE}[1]{}

        \newcommand{\COMMENT}[1]{}
        \newcommand{\TCOMMENT}[1]{}
        
        \newcommand{\REM}[1]{}
}	
{

\usepackage[notcite,notref]{showkeys}
\definecolor{brown}{rgb}{0.6,0.2,0.2}

 \newcommand{\COMMENT}[1]{{{}\uuline{#1}{}}} 
 \newcommand{\DELETE}[1]{{\color{brown}\cancel{#1}{}}}

 \newcommand{\TCOMMENT}[1]{\COMMENT{\parbox{5em}{\tiny #1}}}

\newcommand{\REM}[1]{\marginpar{\bfseries\tiny{{}#1}}}

}

\begin{document}
\begin{sloppypar}
\allowdisplaybreaks
\maketitle
\

\centerline{\scshape Tom\'a\v s Roub\'\i\v cek}
\medskip
{\footnotesize
 \centerline{Mathematical Institute, Charles University} 
\centerline{Sokolovsk\'a 83, CZ-186~75~Praha~8, Czech Republic}
\centerline{and}
\centerline{Institute of Thermomechanics of the ASCR,}
\centerline{Dolej\v skova 5, CZ-182 00 Praha 8, Czech Republic}
\centerline{\emaila}
} % Do not forget to end the {\footnotesize by the sign }

\medskip

\centerline{\scshape Giuseppe Tomassetti}
\medskip
{\footnotesize
 % please put the address of the second  and third author
 \centerline{Universit\`a degli Studi di Roma ``Tor Vergata''}
   \centerline{Dipartimento di Ingegneria Civile e Ingegneria Informatica}
   \centerline{Via Politecnico 1, 00133 Roma, Italy}
\centerline{\emailb}
}

\bigskip 

 %\centerline{(Communicated by the associate editor name)}

\begin{abstract}
A thermodynamically consistent mathematical model for hydrogen adsorption in 
metal hydrides is proposed. Beside hydrogen diffusion, the model accounts for 
phase transformation accompanied by hysteresis, {swelling}, 
temperature and heat transfer, strain, and stress.
We prove existence of solutions of the 
ensuing system of partial differential equations by a carefully-designed, 
semi-implicit approximation scheme. A generalization for
a drift-diffusion of multi-component ionized 
{``gas''} is outlined, too.
\end{abstract}

\section{Introduction}

Hydrogen can be produced from a variety of renewable resources
or in modern 4th-generation nuclear reactors operating at high 
temperatures where hydrogen production by {}water {}hydrolysis advantageously 
serves also to their cooling during periods when electricity cannot be 
produced. Then it is utilized in high-efficiency power generation systems 
with no emission of pollutants based on thermo-chemistry (burning directly hydrogen) or electro-chemistry (using fuel cells, cf.~Section~\ref{sec-fuel-cells} for little more details). Hydrogen contains more energy per unit mass 
than any other available fuel. However, being the lightest element of the 
Periodic Table, it is highly volatile. Thus, in order to be compactly 
stored, standardly it is
compressed in heavy high-pressure tanks or liquefied with recourse to 
expensive cryogenic systems. The lack of an efficient and economical way to 
store hydrogen is the major barrier to the massive commercial implementation of hydrogen-based technologies, especially in the automotive sector \cite{Edwards2008}. A promising alternative to cryogenic and high-pressure hydrogen storage option is provided by solid-state storage, a technology which exploits the property of certain metals and alloys to accommodate hydrogen atoms in their interstitial sites \cite{Libowitz1994}. We propose a mathematical model for hydrogen adsorption in metals. Beside diffusion, the model accounts for phase transformation, 
temperature, strain, and hysteresis, cf.\ e.g.\ \cite{DrClGu??HCHS}. 
Thus, our model, {based on a conventional 
rational mechanics (cf.\ Remark~\ref{rem-RM} below)}, 
extends those proposed and analyzed in 
\cite{Bonetti2012,Bonetti2007,Chiodaroli2011}, 

Since the modeling is entirely new, a detailed derivation is presented in 
the next Section~\ref{sec-model} where also the model is a bit reformulated to 
facilitate mathematical analysis; {of course, various simplifications had 
to been adopted, cf.\ Remark~\ref{rem-simplifications} below.} Mathematical 
results as 
far as existence of weak solutions are summarized in Section~\ref{sec-results} 
while their proof by a carefully designed semi-implicit discretisation in time
is done in Section~\ref{sec-discrete}.
Eventually, in Section~\ref{sec-fuel-cells}, we briefly sketch the 
augmentation of the model for a multicomponent, charged 
(i.e.\ ionized) chemically reacting medium instead of mere 
single-component electro-neutral hydrogen, having in mind 
e.g.\ application to the mentioned fuel cells or to elastic semiconductors.

\section{Model derivation}\label{sec-model}
We consider a solid body, which we identify with a 
{domain} $\Omega$ of the 
three-dimensional space. We regard $\Omega$ as a platform for several 
mutually interacting processes and phenomena affecting the 
kinetics of hydrogen adsorption/desorption \cite{Latroche2004,Libowitz1994}:
\begin{itemize} 
\item \emph{Phase transformation}: at a low concentration, hydrogen atoms form 
a dilute interstitial solid-solution
(\emph{$\alpha$-phase}). 
Increasing the hydrogen concentration causes parts of the solid solution to 
precipitate into a \emph{$\beta$-phase} of larger interstitial concentration 
and lower density. Further addition of hydrogen takes place at constant 
pressure, until the metal is entirely converted into hydride. 
\item \emph{Temperature 
variation}: hydrogenation is 
\emph{exothermic} and \emph{reversible}: when the metal is exposed to 
hydrogen, heat is generated; conversely, heating the hydride drives the 
reaction in the reverse direction. 
\item \emph{Strain and stress}: hydrogenation is accompanied by large 
expansion of the unit cell volume of the crystal. 
{Within this ``swelling''}, volume changes between 
the two phases can vary from 8\% to 30\% and it may 
cause large stresses.
\item \emph{Spatial distribution and transport}: in addition, 
an important feature is distri\-buted-parameter character of such storage 
devices. In particular, the motion of H atoms after 
dissociation of the $\rm H_2$ molecule on the surface is diffusion driven
by gradient of chemical potential, and heat transfer and force equilibrium
must be properly counted.
\end{itemize}
In order to describe the above-mentioned processes we introduce the following 
time-dependent fields on $\Omega$, which we refer to as \emph{primary fields}:
\begin{itemize} 
\item $\bfu$, the displacement field;
\item {$\bbc$, the microstructural phase field;}
\item $\CHI$, the concentration of moles of hydrogen per unit volume;
\item $\vartheta$, the temperature field;
\end{itemize}
The microstructural field is a collection of scalar variables which contains 
information concerning phase transformation and damage. 

We now derive a system of partial differential equations ruling the evolution 
of the primary fields. We do this in two steps.

\emph{Step 1: Balance laws.}
We invoke certain well-accepted thermomechanical prin\-ciples, whose 
statement requires the introduction of some \emph{auxiliary fields}:
\begin{center}
\ \ \ \ \ \ \ \begin{minipage}[t]{0.36\linewidth}

${\boldsymbol\sigma}$ stress,

$\bff$ bulk force,

$\bff_{\rm s}$ surface force,

$\bbs$ internal microforce,

$\bbS$ microstress,

$\bbf$ bulk microforce,

$\bbf_{\rm s}$ external surface microforce,

$e$ internal energy,

\end{minipage}
\hfill 
\begin{minipage}[t]{0.52\linewidth}

$\bfeps(\bfu)=\frac12\left(\nabla\bfu+\nabla\bfu^\top\right)$ small-strain tensor,

$\psi$ free energy,

$\mu$ chemical potential,

$\bfh$ hydrogen flux,

$h$ bulk hydrogen supply,

$h_{\rm s}$ surface hydrogen supply, 

$\bfq$ heat flux,

$q$ bulk heat supply.

\end{minipage}
\end{center}

\medskip

\noindent Each particular specification of space-time evolution of primary 
and auxiliary fields constitutes a \emph{dynamical process}. We require that 
every dynamical process comply with the following balance equations: 
\begin{subequations}\label{balance}
\begin{align}
&\varrho\DDT\bfu-\textrm{div}{\bfsigma}=\bff,\label{paperino}\\
&\bbs-\textrm{div}\bbS=\bbf,\label{qui}\\
&\DT\CHI+\textrm{div}\mathbf h=h,\label{quo}\\
&\DT e+\textrm{div}\mathbf q=q+{\bfsigma}{:}\bfeps(\DT\bfu)
+\bbs{\cdot}\DT\bbm+\bbS{:}\nabla\DT\bbm+\DT\CHI\mu-\mathbf h{\cdot}\nabla
\mu,
\label{qua}
\end{align}
\end{subequations}
where the dot denotes the time derivative.
The statements contained in \eqref{balance} are, in the order: 
the \emph{standard-force balance}, the \emph{microforce balance}, 
the \emph{balance of mass for hydrogen}, and the \emph{balance of internal 
energy}. The corresponding natural conditions on $\partial\Omega$ 
are:\\\vspace*{-2em}
\begin{subequations}\label{boundary}
\begin{align}
&{\bfsigma}\bfn=\bff_{\rm s},\\
&\bbS\bfn=\bbf_{\rm s},\\
&\bfh\cdot\bfn=h_{\rm s},\\
&\bfq\cdot\bfn=q_{\rm s}.
\end{align}
\end{subequations}

Although the number of balance equations equals that of primary fields, the system \eqref{balance} and \eqref{boundary} is under-determined. Such indeterminacy reflects the fact that these laws are common to a wide spectrum of thermomechanical systems. Thus, they cannot single out the particular mathematical model that best fits the system under investigation. One needs indeed additional conditions which can distinguish one particular material from another. These are called \emph{constitutive prescriptions}.\medskip

\emph{Step 2: Second law.} 
A constitutive prescription is typically expressed as a relation between the instantaneous value of a secondary field at a given point and that of a so-called \emph{constitutive list}, a list of quantities obtained by taking the values of primary and secondary fields, or their space/time derivatives. A basic principle that guides the formulation of constitutive prescriptions is the {}requirement that every conceivable dynamical process be consistent with {}the \emph{entropy inequality}:
\begin{equation}\label{pippo}
\DT s\ge \textrm{div}\Big(\frac {\mathbf q}{\vartheta}\Big)+\frac q \vartheta,
\end{equation}
irrespectively of the practical difficulties involved in realizing such a process. Thus, unlike the balance laws \eqref{balance}, the imbalance
\eqref{pippo} is not explicitly stated in the mathematical model, but it is implicitly enforced through a suitable choice of  constitutive prescriptions. 

The entropy inequality is best exploited by replacing, in the list of fields to be specified constitutively, the internal energy with the {}\emph{free energy}:{}
\begin{equation}\label{psi=e-s.theta}
\psi=e-s\vartheta.
\end{equation}
Rewriting \eqref{qua} in terms of $\psi$ and $s$, and substituting 
it into \eqref{pippo}, one arrives at:
\[
\DT\psi+s\DT\vartheta-\mu\DT\CHI\le {\bfsigma}:\DT\bfeps+\bbs\cdot\DT\bbm+\bbS:\nabla\DT\bbm-\mathbf h\cdot\nabla\mu
-\frac 1 \vartheta\mathbf q\cdot\nabla\vartheta,
\]
where we have used the shorthand notation $\bfeps=\bfeps(\bfu)$. A {}standard argument 
due to Coleman and Noll \cite{Coleman1963} {}allows us to conclude that the free energy may 
depend at most on $(\bfeps,\bbm,\nabla\bbm,\CHI,\vartheta)$:
\[
\psi=\varphi(\bfeps,\bbm,\nabla\bbm,\CHI,\vartheta).
\]
Moreover, if one assumes that entropy and chemical potential depend on the same list, 
one obtains 
\[
s=-\partial_\vartheta\varphi,\qquad \mu=\partial_\CHI\varphi.
\]
The dissipation inequality can further be written in a more compact form by 
introducing the splitting:\\\vspace*{-2em}
\begin{align}
&{\boldsymbol\sigma}=\partial_{\bfeps}\varphi+{\bfsigma}^{\rm d},\\
&{\bbs}=\partial_{\bbm}\varphi+\bbs^{\rm d},\\
&{\bbS}=\partial_{\nabla\bbm}\varphi+\bbS^{\rm d}.
\end{align}
With that splitting, one indeed obtains:
\begin{equation}\label{zg}
0\le {\bfsigma}^{\rm d}:\DT\bfeps+\bbs^{\rm d}\cdot\DT\bbm+\bbS^{\rm d}:\nabla\DT\bbm-\bfh\cdot\nabla\mu
-\frac 1 \vartheta\bfq\cdot\nabla\vartheta.
\end{equation}

\emph{Step 3: Constitutive equations.}
To facilitate mathematical analysis but still capturing desired features,
we restrict our attention to the following special constitutive 
ansatz:
\begin{equation}\label{ansatz1}
\varphi(\bfeps,\bbm,\nabla\bbm,\CHI,\vartheta)={\varphi_1}(\bbm,\CHI)+{\varphi_2}(\bfeps,\bbm)+
{\varphi_3}(\bbm,\vartheta)+\vartheta\varphi_4(\bbm,\bfeps)+
\frac\lambda2|\nabla\bbm|^2,
\end{equation}
where $\lambda>0$ is a length-scale parameter.
This ansatz
ensures, e.g., the heat capacity independent of the 
variables whose gradient is not directly controlled, i.e.\ 
$\bfeps$ and $\CHI$, and also the chemical potential independent
of $\bfeps$ and $\vartheta$.

The constitutive equations for entropy and chemical potential are
\begin{subequations}
\begin{align}
&s=-\partial_\vartheta{\varphi_3}(\bbm,\vartheta)-\varphi_4(\bfeps,\bbm),\\
&\mu=\partial_\CHI{\varphi_1}(\bbm,\CHI).
\end{align}
\end{subequations}
On defining {$\omega:=\varphi-\vartheta\partial_\vartheta\varphi$, in view 
of \eqref{ansatz1} we have}
\begin{align}\label{def-of-omega}
\omega(\bbm,\vartheta)=
\varphi_3(\bbm,\vartheta)-\vartheta
\partial_\vartheta\varphi_3(\bbm,\vartheta),
\end{align}
the constitutive equation  for internal energy 
{$e=\psi+s\vartheta$, cf.\ \eqref{psi=e-s.theta},}
is
\[
e=\varphi_1(\bbm,\CHI)+\varphi_2(\bfeps,\bbm)+\omega(\bbm,\vartheta)
+\frac\lambda2|\nabla\bbm|^2.
\]
As constitutive prescriptions for the dissipative parts of the auxiliary fields we choose\\[-1.2em]
\begin{subequations}\label{interm}
\begin{align}
&{\boldsymbol\sigma}^{\rm d}=\mathbb D\bfeps(\DT\bfu),\\
&\bbs^{\rm d}\in \alpha\DT\bbm+\partial\zeta(\DT\bbm),\label{dissip-pot-m}\\
&\bbS^{\rm d}=0,\\
&\bfq=-\bbK(\bfeps,\bbm,\CHI,\vartheta)\nabla\vartheta,\label{aaa}\\
&\bfh=-\bbM(\bfeps,\bbm,\CHI,\vartheta)\nabla\mu.\label{bbb}
\end{align}
\end{subequations}
Here $\mathbb D$ is a 2nd-order positive-definite viscosity-moduli 
tensor,  $\alpha>0$ counts for rate effects in
 evolution of $\bbm$,  $\zeta$ is a 
convex function homogeneous 
of degree one; note that $\zeta$ is typically nonsmooth at 0, which
counts for activation of evolution of $\bbm$. Moreover, 
$\bbK$ and $\bbM$ are respectively 2nd-order positive-definite
heat-conductivity and hydrogen-diffusivity tensors. 
We also eventually set $h=0$ and $\bbf=0$. We therefore arrive at the 
following system:
\begin{subequations}\label{orig-syst}
\begin{align}
&\varrho\DDT\bfu-
\mathrm{div}\big(
\partial_{\bfeps}\varphi_2(\bfeps(\bfu),\bbm)
+{\vartheta\partial_{\bfeps}\varphi_4(\bfeps(\bfu),\bbm)}
+\mathbb D\bfeps(\DT{\bfu})\big)=\bff,\label{balmech-}\\
&\alpha\DT\bbm-\lambda\,\Delta \bbm
+\partial_\bbm{\varphi_1}(\bbm,\CHI)+\partial_\bbm{\varphi_2}(\bfeps(\bfu),\bbm)
\nonumber\\
&\qquad\qquad\qquad\qquad\ 
+\partial_\bbm{\varphi_3}(\bbm,\vartheta)
+{\vartheta\partial_{\bbm}\varphi_4(\bfeps(\bfu),\bbm)}
%{}\in-\partial\zeta(\dot\chi),
+\partial\zeta(\dot\chi)\ni0,\label{phasetransf}
\\&
\DT\CHI-\mathrm{div}\big(\bbM(\bfeps(\bfu),\bbm,\CHI,\vartheta)\nabla\mu\big)=0,
\label{EEd-t2-}
\\
&\DT w-\mathrm{div}\Big(\bbK(\bfeps(\bfu),\bbm,\CHI,\vartheta)\nabla
\vartheta\Big)=
\big(\mathbb D\bfeps(\DT\bfu)
+{\vartheta\partial_{\bfeps}\varphi_4(\bfeps(\bfu),\bbm)}\big)
{:}\bfeps(\DT\bfu)\nonumber
\\&\null\qquad\qquad\qquad\qquad\
+\big(\alpha\DT\bbm+\partial_{\bbm}{\varphi_3}(\bbm,\vartheta)
+{\vartheta\partial_{\bbm}\varphi_4(\bfeps(\bfu),\bbm)}
\big){\cdot}\DT\bbm\nonumber
\\&\null\qquad\qquad\qquad\qquad\
+\zeta(\DT\bbm)+\bbM(\bfeps(\bfu),\mathbbmss m,\CHI,\vartheta)
\nabla\mu{\cdot}\nabla\mu+q,
\\\label{def-of-mu}
&\mu=\partial_\CHI{\varphi_1}(\bbm,\CHI),
\\
&w=\omega(\bbm,\vartheta),
\end{align}
\end{subequations}
{where $\omega$ is from \eqref{def-of-omega}.}
We make the following {natural} assumption {which, in fact, says 
positivity of the heat capacity}:
\begin{align}
&\partial_\vartheta 
\omega={-\vartheta\partial^2_{\vartheta\vartheta}\varphi}=
-\vartheta\partial^2_{\vartheta\vartheta}\varphi_3>0.
\label{heatc}
\end{align}
Then, the inverse to $\omega(\bbm,\cdot)$ does exist and
we can express $\vartheta$ as 
\[
\vartheta={[\omega(\bbm,\cdot)]^{-1}(w)=:}\theta(\bbm,w),
\]
which allows us to eliminate temperature $\vartheta$ from the system 
\eqref{orig-syst}. {} The symbol  $\ni$ appearing in formula \eqref{phasetransf} (and in formulas \eqref{gilbmod-t}, \eqref{TGMd-2}, and \eqref{TGMd-2-comp} below) means that the right--hand side is included in the left--hand side, which is a set, since $\zeta$ and $\varphi_2$ are non--smooth (see also \eqref{def-of-phi2} below).
{}

Moreover, we will be a bit more specific in \eqref{ansatz1}.
A typical contribution to the free energy is
\begin{align}\label{basic-ansatz}
{\frac12\bbC(\bfeps{-}\bfepstr\bbc{-}\vartheta\bfalpha)
{:}(\bfeps{-}\bfepstr\bbc{-}\vartheta\bfalpha)
+\frac k2\big|\bbc{-}a(\CHI)\big|^2+\phi_1(\CHI)+\phi_3(\vartheta)
+\delta_K(\bbc)}
\end{align}
where  
$\bbC$ is a 4th-order {elastic-moduli} 
tensor, 
$\bfepstr$ is a 2nd-order tensor which incorporates the effect of dilation due 
to the microstructural parameter $\bbc$ within metal/hydride phase 
transformation, $\bfalpha$ is a  
tensor accounting for thermal dilation, $\phi_1$ and $\phi_3$
are the simplest variant of the contribution to the {chemical potential}
and the heat capacity, respectively, and 
{$\delta_K$ is an indicator function of a convex set $K\subset\R^N$
from which the phase-field $\bbc$ is assumed to range. Assuming $\bbm$ is 
scalar-valued, $\bfepstr$ the unit matrix, and $k$ large,
we get essentially an isotropic {\it swelling} controlled nonlinearly 
by the hydrogen concentration by $\bbc\sim a(\CHI)$
while allowing still $\varphi_1(\bbm,\cdot)$ to be
uniformly convex, as needed later in \eqref{est-of-nabla-concentration}.}
Obviously, {in view of \eqref{ansatz1}, the specific choice}
\eqref{basic-ansatz} 
leads to
\begin{subequations}
\begin{align}\label{def-of-phi1}
&{\varphi_1(\bbm,\CHI)=\frac k2\big|\bbc{-}a(\CHI)\big|^2+\phi_1(\CHI),}
\\\label{def-of-phi2}
&{\varphi_2}(\bfeps,\bbm)
=\frac12\bbC
(\bfeps{-}\bfepstr\bbc){:}(\bfeps{-}\bfepstr\bbc)+\delta_K(\bbc)
,\\\label{def-of-phi3}
&
\varphi_3(\bbm,\vartheta)=
\frac12\vartheta^2\bbC\bfalpha{:}\bfalpha
+\vartheta\bbC\bfalpha{:}\bfepstr\bbc+\phi_3(\vartheta),
\\
&\varphi_4(\bfeps,\bbm)=
-\bbC\bfalpha{:}\bfeps.
\end{align}
\end{subequations}
Note that, {in \eqref{def-of-phi3}, 
$\partial_{\vartheta\vartheta\bbm}^3\varphi_3\equiv0$,} 
which makes the heat capacity independent of $\bbm$, but we can consider
more generally the contribution $\phi_3$ dependent also on $\bbm$ to 
reflect different heat capacity of metal and of hydride, and therefore 
we do not restrict ourselves to a particular form of $\varphi_3$ 
in \eqref{ansatz1} 
but make only certain technical assumptions below, {cf.\ \eqref{q2}.} 
Similarly, we keep the treatment of $\varphi_1$ in a nonspecified way. 
{In fact, the specific form \eqref{def-of-phi2} is a simplified
linearization of the so-called St.Venant-Kirchhoff potential
but, when derived from the St.Venant-Kirchhoff
form by linearizing the stress response, it results still in some
other terms, cf.\ \cite[Sect.\,5.4]{DreDud??MSIG}, which here
was neglected rather for notational simplicity.}

Thus, on setting\\[-1.3em] 
\begin{subequations}
\begin{align}
&\bfsigma_{\rm a}(\bbm,w):=-\theta(\bbm,w)\bbC\bfalpha,\\
&{\bbs_{\rm a}}(\bbm,w):=\partial_\bbm{\varphi_3}(\bbm,\theta(\bbm,w))
,
\\
&\mathsf K(\bfeps,\bbm,\CHI,w):=\bbK(\bfeps,\bbm,\CHI,\theta(\bbm,w)),\\
&\mathsf{L}(\bfeps,\bbm,\CHI,w):=
\bbK(\bfeps,\bbm,\CHI,\theta(\bbm,w))\otimes\partial_\bbm\theta(\bbm,w),\\
&\mathsf M(\bfeps,\bbm,\CHI,w):=\bbM(\bfeps,\bbm,\CHI,\theta(\bbm,w)),
\end{align}
\end{subequations}
the original system \eqref{orig-syst} is transformed into
\begin{subequations}\label{BVP-t}
 \begin{align}\label{balmech}
&\!\!\varrho\DDT\bfu-
\mathrm{div}\big(
{\bbC(\bfeps(\bfu){-}\bfepstr\bbm)}
+{\bfsigma_{\rm a}}(\bbm,w)+\mathbb D\bfeps(\DT\bfu)\big)=\bff,
\\&\!\!
%{}
\alpha\DT\bbm+\partial\zeta(\DT\bbm)-\lambda\,\Delta \bbm\!
\label{gilbmod-t}
+\partial_\bbm{\varphi_1}(\bbm,\CHI){+}
{\bfepstr^\top\bbC(\bfepstr\bbm{-}\bfeps(\bfu))}
{+}{\bbs_{\rm a}}(\bbm,w){+}N_K^{}(\bbm)
\ni0,
%\in -\partial\zeta(\DT\bbm){-}N_K^{}(\bbm),
\\
&\!\!\DT\CHI-\mathrm{div}\big(\mathsf M(\bfeps(\bfu),\bbm,\CHI,w)\nabla\mu\big)=0,
\label{EEd-t2}
\\&\!\!\DT w-\mathrm{div}\big(\mathsf K(\bfeps(\bfu),\bbm,\CHI,w)
\nabla w{+}\mathsf{L}(\bfeps(\bfu),\bbm,\CHI,w)
\nabla\bbm\big)=\big({\bfsigma_{\rm a}}(\bbm,w){+}
\mathbbm D\bfeps(\DT\bfu)\big){:}\bfeps(\DT{\bfu})
\nonumber
\\&\qquad\qquad
+\big(\bbs_{\rm a}(\bbm,w)+\alpha\DT\bbm\big){\cdot}\DT\bbm
+\zeta(\DT\bbm)
+\mathsf M(\bfeps(\bfu),\mathbbmss m,\CHI,w)\nabla \mu{\cdot}\nabla\mu+q,
\label{heatequation}
\\&\!\mu=\partial_\CHI{\varphi_1}(\bbm,\CHI),\label{chempot}
\end{align}
\end{subequations}
{where $N_K^{}=\partial\delta_K$ denotes standardly the normal cone to 
the convex set $K$.}
The boundary conditions {\eqref{boundary} now} take the form:
\begin{subequations}\label{bcc}
 \begin{align}
&\big(
\bbC(\bfeps(\bfu){-}\bfepstr\bbm)
+{\bfsigma_{\rm a}}(\bbm,w)+\mathbb D\bfeps(\DT\bfu)\big)\mathbf n
=\bff_{\rm s},
\label{BC-t-1--}
\\
&
\,
\DELETE{\mathsf M(\bfeps(\bfu),\bbm,\CHI,w)}
\nabla{\bbm}
\cdot{\mathbf n}=0,
\\
&\label{bc34}
\,\mathsf M(\bfeps(\bfu),\bbm,\CHI,w)\nabla \mu
\cdot\mathbf n=h_s,
\\
&\label{BC-t-2}
\big(\mathsf K(\bfeps(\bfu),\bbm,\CHI,w)
\nabla w+\mathsf{L}(\bfeps(\bfu),\bbm,\CHI,w)
\nabla{\bbm}\big)\cdot\mathbf n=q_{\rm s}.\end{align}
Using the convention like $\bfu(\cdot,t)=:\bfu(t)$,
we complete the system by the initial con\-ditions:\\[-1.3em]
\begin{align}
&\label{IC}
\bfu(0)=\bfu_0,\qquad\DT\bfu(0)=\bfv_0,
\qquad\bbm(0)=\bbm_0,\qquad\CHI(0)=\CHI_0,\qquad w(0)=w_0,
\end{align}
\end{subequations}
where we have set $w_0=\omega(\bbm_0,\vartheta_0).$
We henceforth shall use the abbreviation for {the so-called stored energy,
i.e.\ the temperature independent part} of the free energy: 
\begin{align}\label{def-of-phi12}
\varphi_{12}(\bfeps,\bbm,\CHI):=
{\varphi_1}(\bbm,\CHI)+{\varphi_2}(\bfeps,\bbm)
\end{align}
By testing 
{(\ref{BVP-t}a,b,c,d)}
respectively with $\DT\bfu$, $\DT\bbm$, $\mu$, and by a constant $\nu$,
integrating by parts {in time, using Green's formula with}
the boundary conditions \eqref{bcc}, and taking into account \eqref{chempot}
so that\\[-1.3em]
\begin{align}\label{fundamental-test}
\DT\CHI\mu=\frac{\partial}{\partial t}\varphi_1(\bbm,\CHI)
-\partial_\bbm\varphi_1(\bbm,\CHI)\DT\bbm,
\end{align} 
we obtain the following identity:
\begin{align}\nonumber
&\int_\Omega \frac\r2\big|\DT\bfu(t)\big|^2
+\varphi_{12}(\bfeps(t),\bbm(t),\CHI(t))+\frac\lambda2|\nabla\bbm(t)|^2
+\nu 
w(t) \,\d x
\\[-.3em]&\nonumber
\quad +(1{-}\nu)\int_0^t\!\!\int_\Omega\!
\bbD\bfeps(\DT\bfu){:}\bfeps(\DT\bfu)+\alpha\big|\DT\bbm\big|^2
{+\zeta\big(\DT\bbm\big)}
+\mathsf M(\bfeps(\bfu),\bbm,\CHI, w)\nabla\mu{\cdot}\nabla\mu\,\d x\d t
\\\nonumber&
=(1{-}\nu)\int_0^t\!\!\int_\Omega
\bfsigma_{\rm a}(\bbm,w){:}\bfeps(\DT\bfu)+\bbs_{\rm a}(\bbm,w){\cdot}\DT\bbm
\,\d x\d t
\\\nonumber &\quad 
+\nu \int_0^t\!\!\int_\Omega q 
\,\d x\d t+\int_0^t\!\bigg(\int_\Omega \bff{\cdot}\DT\bfu\,\d x
+\int_\Gamma{\bff_{\rm s}}{\cdot}\DT\bfu
+{h_{\rm s}}{\cdot}\DT\mu+{\nu} {q_{\rm s}}\,\d S
\bigg)\d t
\\&\quad
+\nu\int_\Omega w_0\,\d x+\int_\Omega\frac\r2|\bfv_0|^2
+\varphi_{12}(\bfeps(\bfu_0),\bbm_0,\CHI_0)+\frac\lambda2|\nabla\bbm_0|^2
\,\d x.\label{balance3} 
\end{align}
For $\nu=0$, the identity \eqref{balance3} represents the mechanical energy 
balance. For $0<\nu<1$, both the internal energy and dissipative terms 
are seen; henceforth,  
a discrete 
version of this estimate will be used in the proof of Lemma \ref{lem-2}
for $\nu=1/2$. 
Eventually, for $\nu=1$, we recover the standard total-energy balance;
note that the dissipative terms (and also adiabatic terms) then vanish.

\begin{remark}\label{rem-simplifications}
Of course, the above model adopted a lot of simplifications of the
actual situations in hydride {}storage{}. In particular 
the concept of small strains may be questionable at some situations,
possible damage is essentially neglected, although formally it can 
be involved in a general form of $\varphi_2$ in \eqref{ansatz1} below
but a lot of analytical considerations seem to be difficult to 
be straightforwardly adapted. Also temperature dependence of the 
chemical potential of hydrogen is neglected, i.e.\ $\phi_1$ in
\eqref{ansatz1} does not depend of $\theta$. Further, the 
chemical reaction in the multi-component system metal/hydrid/hydrogen is 
basically neglected and hydride is modeled as a mixture of metal and hydrogen 
with essentially the possibility to obtain the same thermomechanical 
response of the phase transformation 
as the corresponding chemical reaction (and, in addition, we can easily model
the activated hysteretic response related with this phase transformation). 
\end{remark}

\begin{remark}\label{rem-RM}
{{}The {}thermodynamics of our model {}follows {}essentially a classical approach
based on rational mechanics and Clausius-Duhem inequality,
cf.\ e.g.\ \cite{Bowe76CP}. There are some variants of this general scenario
\cite{DegMaz84NET,LeJoCa08UNET,KjeBed08NETH,Mull85T} which are to some extent 
equivalent under some simplifications like those in 
Remark~\ref{rem-simplifications} above.
}

\end{remark}

\section{Weak solutions and their existence}\label{sec-results}

\def\deltauu{\frac{\bfu\kt-\bfu\kkt}\tau}
\def\deltau{\dtt\kt\bfu}
\def\deltauuu{\dtt_\tau^{k-1}\bfu}
\def\deltam{\dtt\kt\bbm}
\def\deltamm{\frac{\bbm\kt-\bbm\kkt}\tau}
\def\deltac{\dtt\kt\bbc}
\def\deltad{\dtt\kt\bbd}
\def\deltaee{\bfeps(\frac{\bfu\kt-\bfu\kkt}\tau)}
\def\deltae{\bfeps(\deltau)}
\def\deltaeee{\dtt\kt\bfeps}
\def\deltaxx{\frac{\CHI^k_\tau-\CHI\kkt}\tau}
\def\deltax{\dtt\kt\CHI}
\def\deltaww{\frac{w^k_\tau-w\kkt}\tau}
\def\deltaw{\dtt\kt w}
\def\vphi{\varphi_{12}}

Let us summarize the qualification on the data on which we will rely during 
the analysis of the initial-boundary-value problem \eqref{BVP-t}.
We confine ourselves to the (physically relevant) three-dimensional case.
We consider a fixed time horizon $T$ and abbreviate $Q:=\Omega{\times}(0,T)$
and $\Sigma:=\Gamma{\times}(0,T)$, with $\Gamma$ a boundary of the 
domain $\Omega\subset\R^3$ assumed Lipschitz.
{In the following, we shall use some classical notation for function spaces,
in particular the Lebesgue spaces $L^p$, the Sobolev spaces $W^{k,p}$
and, in particular, $H^k=W^{k,2}$, and vector-valued functions.} 
We 
suppose that 
\begin{subequations}\label{DQ}
\begin{align}\nonumber
&\bbC,\ \mathbb D,\ \partial^2_{\CHI\CHI}{\varphi_1}(\bbm,\CHI),\ 
 \mathsf M(\bfeps,\bbm,\CHI,w),\ 
 \mathsf K(\bfeps,\bbm,\CHI,w)
\\&\hspace*{12em}\text{are (uniformly) positive definite},\label{pos1}
\\\label{growth-of-phi1}
&\max\big(|\varphi_1(\bbm,\CHI)|,
|\partial_\bbm\varphi_1(\bbm,\CHI)|\big)\le C\big(1
+|\CHI|^3\big),
\\&|\mathsf M(\bfeps,\bbm,\CHI,w)\partial^2_{\CHI\CHI}\varphi_1(\bbm,\CHI)| \le C(1+|\CHI|^{6-\epsilon}),\label{growth1}
 \\\label{semiconvexity-phi0}
 &\varphi_1\in {\rm C}^2(K{\times}\R^+;\R),\ \ 
\partial^2_{\bbm\bbm}\varphi_1(\bbm,\CHI)\ \  \text{ is bounded from below},
\\\label{bbg}
&\partial^2_{\bbm\CHI}\varphi_1(\bbm,\CHI)\ \ \text{is bounded, and }\
{ \partial^2_{\bbm\CHI}\varphi_1(\bbm,\CHI)=0\ \text{ for }\CHI=0,}
\\\label{Mbound}
&\mathsf M(\bfeps,\bbm,\CHI,w)\ \ \text{is bounded},
 \\\label{phi0-coerc}
 &\exists\epsilon>0:\qquad\varphi_1(\bbm,\CHI)\ge
{\epsilon|\CHI|^2},
 \ \ \ \ \varphi_1(\bbm,\cdot)\text{ convex},
 %\\&{\exists K\subset\R^N\text{ bounded}\ \forall\bbm\in\R^N
 %{\setminus}K:
 %%\ \ \bbC'(\bbm)=0,\ 
 %\bfepstr'(\bbm)=0,\ \varphi_3(\bbm,\CHI)=0,}
 %\\\label{L2-rhs}
 %&{\text{growth so that }\partial_\bbm{\varphi_1}(\bbm,\CHI)
 %+\partial_\bbm{\varphi_2}(\bfeps(\bfu),\bbm)+{\bbs_{\rm a}}(\bbm,w)\in L^2(Q;\R^N).\COMMENT{TO\ MODIFY!!!!}}
 \\
 &
\zeta:\R^N\to\R\text{ convex 1-homogeneous.}
\\&{K\subset\R^N\ \text{ bounded, convex, closed,}}
\\\label{q1}
&{\bfsigma_{\rm a}\in C(K{\times}\R^+;\R^N),}\qquad\
|\bfsigma_{\rm a}(\bbm,w)|\le C\sqrt{1+\varphi_1(\bbm,\CHI)+w},\\
\label{q2}
&{\bbs_{\rm a}\in C(K{\times}\R^+;\R^{3\times3}),}\qquad
|\bbs_{\rm a}(\bbm,w)|\le C\sqrt{1+\varphi_1(\bbm,\CHI)+w},\\
\label{q3}
&|\mathsf L(\bfeps,\bbm,\CHI,w)|\le C\sqrt{1+w}.  
\end{align}
{}Moreover, 
we need the 
qualification of the right-hand sides and the initial data:
\begin{align}
&\bff\!\in\!L^2(I;L^{6/5}(\Omega;\R^3)),\quad\bff_{\rm s}\!\in\!L^2(I;L^{4/3}(\Gamma;\R^3)),
\quad q\!\in\!L^1(\Omega),
\quad q_{\rm s}\!\in\!L^1(\Gamma),
\\\label{IC-ass-1}
&
\bfu_0\!\in\!H^1(\Omega;\R^3),\ \ \bfv_0\!\in\!L^2(\Omega;\R^3),\ \ 
\bbm_0\!\in\!H^1(\Omega;\R^N),\ \ \ {\bbm_0\!\in\!K\text{ a.e.\ on }\Omega,}
\\\label{IC-ass-2}
&\CHI_0\!\in\!H^1(\Omega),\ \ \ {\CHI_0\ge 0,}\ \ \ 
w_0\!\in\!L^1(\Omega),\ \ \ {w_0\ge 0}.
\end{align}
\end{subequations}
{We note that \eqref{q3} will be used in the derivation of the 
estimate on $\nabla w$, see \eqref{est-of-nabla-w} below. Also note that 
(\ref{DQ}a,d,e)
is not in conflict with the example 
\eqref{def-of-phi1} where 
$\partial^2_{\bbm\CHI}\varphi_1(\bbm,\CHI)=-ka'(\CHI)$
and $\partial^2_{\CHI\CHI}\varphi_1(\bbm,\CHI)=k(a(\CHI){-}\bbm)a''(\CHI)
+\phi_1''(\CHI)+ka'(\CHI)^2$, 
while $\partial^2_{\bbm\bbm}\varphi_1(\bbm,\CHI)=k$
so that \eqref{pos1} 
needs $k(a(\CHI){-}\bbm)a''(\CHI)
+\phi_1''(\CHI)+ka'(\CHI)^2\ge\epsilon>0$,
\eqref{bbg} needs $a'$ bounded with $a'(0)=0$ 
while \eqref{semiconvexity-phi0} is here satisfied automatically.
}

\begin{definition}[{\sc Weak solutions}]\label{def}
We say that the 
six-tuple $(\bfu,\bbm,\CHI,w,\mu,\xi)$ 
with
\begin{subequations}\begin{align} 
&\bfu\in H^1(I;H^1(\Omega;\R^3)),\quad
\\&\bbm\in L^\infty(I;H^1(\Omega;\R^N))
\cap{H^1(I;L^2(\Omega;\R^N))},\quad
\\&\CHI\in L^\infty(I;L^2(\Omega)),
\\&w\in L^\infty(I;L^1(\Omega))\cap L^1(I;W^{1,1}(\Omega)),\quad
\\&\mu\in L^\infty(I;H^1(\Omega)),
\\&\xi\in L^2(Q;\R^N),\qquad \xi\in N_K^{}(\bbm)\ \ \text{ a.e.\ on }\ Q,
\end{align}\end{subequations}
is a weak solution to the 
initial-boundary-value problem \eqref{BVP-t}--\eqref{bcc}
if
\begin{subequations}
\begin{align}\nonumber
\int_Q\!
\big(
\bbC(\bfeps(\bfu){-}\bfepstr\bbm)
+\bfsigma_{\rm a}(\bbm,w)+\mathbb D\bfeps(\DT\bfu)\big){:}\bfeps(\bfz)
-\varrho\DT\bfu{\cdot}
\DT\bfz\,\d x\d t\qquad\qquad\qquad\quad
\\[-.8em]
=\int_Q\!\bff{\cdot}\bfz\,\d x\d t
+\int_\Sigma\!\bff_{\rm s}{\cdot}\bfz\,\d S\d t
+\int_\Omega\!\bfv_0{\cdot}\bfz(0)\,\d x
\end{align}
for any $\bfz\in C^1(\overline Q;\R^3)$ such that $\bfz(T)=0$,
\begin{align}\nonumber
&\int_Q\!\zeta(\bbv)+
\big(\alpha\DT\bbm+
\partial_\bbm{\varphi_1}(\bbm,\CHI)
+{\bfepstr^\top\bbC(\bfepstr\bbm{-}\bfeps(\bfu))}
+\bbs_{\rm a}(\bbm,\CHI)
+{\xi}
\big){\cdot}(\bbv{-}\DT\bbm)
\\[-.7em]&\label{def-of-m}\qquad
+\lambda\nabla\bbm{:}
\nabla\bbv
\,\d x\d t+\int_\Omega\!\frac\lambda2|\nabla\bbm_0|^2\,\d x
\ge\int_Q
\zeta(\DT\bbm)\,\d x\d t+\int_\Omega\!\frac\lambda2|\nabla\bbm(T)|^2\,\d x
\end{align}
for any  $\bbv\in C^1(\overline Q;\R^N)$,
\begin{align}\label{def-of-chi}
\int_Q\!\bbM(\bfeps(\bfu),\bbm,\CHI,\vartheta)\nabla\mu{\cdot}
\nabla v-\CHI\DT v\,\d x\d t=\int_\Sigma h_sv\,\d S\d t+\int_\Omega\CHI_0v\,\d x
\end{align}
for all $v\in C^1(\overline Q)$ with $v(T)=0$,
\begin{align}\nonumber
&\int_Q \Big(\mathsf K(\bfeps(\bfu),\bbm,\CHI,w)\nabla w+\mathsf L(\bfeps(\bfu),\bbm,\CHI,w)\nabla\bbm\Big)\cdot\nabla w-w\dot v\, \d x\d t
\\[-.7em]
&\quad=\int_\Omega w_0 v(0)\,\d x+\int_\Sigma q_s v\, \d S \d t+\int_Q \Big(q+\big(\sigma_{\rm a}(\bbm,w)+\mathbb D\bfeps(\dot\bfu)\big){:}\bfeps(\dot\bfu)
\nonumber
\\[-.6em]
&\quad\ \ \ \ \ \ \ \ \ 
+\big(\bbs_{\rm a}(\bbm,w)+\alpha\dot\bbm\big){\cdot}\dot\bbm+\zeta(\dot\bbm)
\mathsf M(\bfeps(\bfu),\bbm,\CHI,w)\nabla\mu{\cdot}\nabla\mu\Big)v\,\d x\d t
\end{align}
for all $v\in C^1(\overline Q)$ such that $v(T)=0$, and eventually
\begin{align}
\mu=\partial_\CHI\varphi_1(\CHI,\bbm)\quad\text{  a.e. in }Q.
\end{align}
\end{subequations}
\end{definition}
The above definition obviously arises from \eqref{BVP-t}--\eqref{bcc}
by using standard concept. The inequality \eqref{def-of-m}
arises by using additionally the identity
\begin{align}\label{by-part-for-m}
\int_Q \Delta\bbm{\cdot}\DT\bbm\,\d x\d t
=\frac12\int_\Omega\big|\nabla\bbm(0)\big|^2-\big|\nabla\bbm(T)\big|^2\,\d x,
\end{align}
which can rigorously be justified if $\Delta\bbm\in L^2(Q;\R^N)$, 
cf.~\cite[Formula (3.69)]{Podio-Guidugli2010}. At this occasion,
let us emphasize that $\nabla\DT\bbm$ is not well defined as a function
on $Q$ so that we avoid using 
$\int_Q\nabla\bbm{:}\nabla(\bbv{-}\DT\bbm)\,\d x\d t$.

\begin{theorem}[{\sc Existence of weak solutions}]\label{thm}
{Let assumptions \eqref{DQ} hold true.} Then 
\eqref{BVP-t}--\eqref{bcc} has at least one weak solution 
$(\bfu,\bbm,\CHI,w,\mu)$ according Definition~\ref{def}. Moreover, { 
\begin{subequations}\label{additional}\begin{align}
&\r\DDT\bfu\in L^2(I;H^1(\O;\R^3)^*),
\\
&\DT w\!\in\!L^1(I;H^3(\O)^*)\ \text{ and }\ w\!\in\!L^r(I;W^{1,r}(\Omega))
\ \text{ for any }1\le r<5/4,
\\
&\DT\CHI\in L^2(I;H^1(\O)^*),
\\\label{additional-1}
&\Delta\bbm\in L^2(Q;\R^N),
\end{align}\end{subequations}
and {this} solution is consistent with the energy conservation equation 
\eqref{balance3} with $\nu=1$, as well as with $\CHI\ge0$ and $w\ge0$ on $Q$. 
If $\Omega$ is smooth, then even $\bbm\in L^2(I;H^2(\Omega;\R^N))$.
}
\end{theorem}

We will prove this theorem in Section~\ref{sec-discrete}  by 
a semi-implicit time discretisation in a more or less 
constructive manner, except the fixed-point argument behind the boundary-value 
sub-problems \eqref{gilbmod-disc}--\eqref{bc34} and
\eqref{heatequation}--\eqref{BC-t-2-2}
and selection of converging subsequences. 
{The additional properties \eqref{additional} follow from 
\eqref{est-of-nabla-w} and \eqref{apriori-II+}. The $H^2$-regularity of 
$\bbm$ is a standard consequence of \eqref{additional-1}.
For more detailed modes of convergence of the mentioned approximate
solutions we refer to \eqref{15.-strong} and \eqref{15.-strong+} below.}

\section{Analysis of \eqref{BVP-t}--\eqref{bcc} by semidiscretisation in time}
\label{sec-discrete}
We will prove existence of a weak solution to the initial-boundary-value problem \eqref{BVP-t} by a carefully constructed semi-implicit discretisation 
in time which, at the same time, will decouple the system to a sequence of convex minimization problems combined with a diffusion equation, and provide thus a rather efficient conceptual numerical strategy. 

In comparison with the fully implicit time discretisation (i.e.\ the so-called Rothe method), our discretisation will allow for a simpler (and constructive) proof
of existence of the discrete solutions, weaker assumptions about convexity mode of the stored energy, but we need to impose a bit stronger growth qualification of the data than required by the nature of the continuous system \eqref{BVP-t}.

We use an equidistant partition of the time interval $I=[0,T]$ with a time 
step $\tau>0$, assuming $T/\tau\in\N$, and denote $\{\bfu\kt\}_{k=0}^{T/\tau}$ 
an approximation of the desired values $\bfu(k\tau)$,
and similarly $\bbm\kt$ is to approximate $\bbm(k\tau)$, etc.
Further, let us abbreviate by $\dtt\kt$ the backward difference
operator, i.e.\ e.g.\  $\deltau:=\deltauu$, and similarly 
also $[\dtt\kt]^2\bfu=\dtt\kt[\deltau]=\frac{{\bfu}\kt{-}
2{\bfu}\kkt{+}{\bfu}_\tau^{k-2}}{\tau^2}$, or 
$\deltam:=\deltamm$, $\deltax:=\deltaxx$, etc. 
Then, using also notation \eqref{def-of-phi12}, 
we devise the following semi-implicit discretisation:
\allowdisplaybreaks[1]
\begin{subequations}\label{TGMd}
 \begin{align}
 &\varrho[\dtt\kt]^2\bfu
 -{\rm div}\Big(\bbC(\bfeps(\bfu\kt){-}\bfepstr\bbc\kt
)
 +{\bfsigma_{\rm a}}({\bbm\kkt},w\kkt)
\label{TGMd-1}
+\mathbb D\bfeps(\deltau)\Big)=\bff\kt,
\\&
%{}
\alpha\deltac
+\partial\zeta\left(\deltac\right)
-\lambda\Delta\bbc\kt
+\partial_\bbc{\varphi_1}(\bbc\kt,\CHI\kkt)
+\bfepstr^\top\bbC(\bfepstr\bbc\kt{-}\bfeps(\bfu\kt))
\nonumber\\&
%{}
\hspace{6em}
+{\bbs_{\rm a}}({\bbm\kkt},w\kkt)+\xi\kt\ni0 \quad\text{ with }
\quad\xi\kt\in N_K^{}(\bbc\kt),
\label{TGMd-2}
 \\&
 \dtt\kt\CHI-\textrm{div}\big(\mathsf M(\bfeps(\bfu\kt),
 \bbm\kt,\CHI\kt,w\kkt)\nabla\mu\kt\big)=0,\label{gilbmod-disc}
 \\
 &
 \dtt\kt w-\mathrm{div}\big(
\mathsf{K}(\bfeps(\bfu\kt),\bbm\kt,\CHI\kt,w\kt)\nabla w\kt
+\mathsf{L}(\bfeps(\bfu\kt),\bbm\kt,\CHI\kt,w\kt)\nabla\bbm\kt
 \big)\nonumber
 \\[-.3em]&\hspace{6em}=q\kt+\big({\bfsigma_{\rm a}}({\bbm\kkt},w\kt)+
\frac{\mathbbm D\bfeps(\deltau)\big){:}\bfeps(\deltau)}
{{1+\tau|\bfeps(\deltau)|^2}}
\nonumber
 \\[-.3em]
 &\hspace{6em}+\big(\bbs_{\rm a}({\bbm\kkt},w\kt)
 +\alpha\deltam\big)\cdot\deltam
 \nonumber
 \\[-.3em]\label{heatequation-disc}&\hspace{6em}+\zeta(\deltam)
+\frac{\mathsf M(\bfeps(\bfu\kt),\bbm\kt,\CHI\kt,w\kkt)
\nabla\mu\kt\cdot\nabla\mu\kt}{1+\tau|\nabla\mu\kt|^2},
 \\
 &\mu\kt=\partial_{\CHI}{\varphi_1}(\bbm\kt,\CHI\kt),
\end{align}
\end{subequations}
for $k=1,...,T/\tau$, together with the boundary conditions
\begin{subequations}\label{BC-t}
\begin{align}
&\big(\bbC(\bfeps(\bfu\kt){-}\bfepstr\bbc\kt)+\bfsigma_{\rm a}({\bbm\kkt},w\kkt)
+\mathbb D\bfeps(\deltau)
\big)\mathbf n=\bff_{\rm s,\tau}^k,
\label{BC-t-1-}
\\\label{BC-TGMd-2}
&
\frac{\partial\bbc\kt}{\partial\mathbf n}=0,
\\
&\label{bc34+}
\mathsf M(\bfeps(\bfu\kt),\bbm\kt,{\CHI\kt},w\kkt)\nabla\mu\kt
\cdot\mathbf n=h_{\rm s,\tau}^k,
\\
&\label{BC-t-2-2}
\big(
\mathsf{K}({{}\bfeps(\bfu\kt),\bbm\kt},\CHI\kt,{w\kt})\nabla w\kt
+\mathsf{L}(\bfeps(\bfu\kt),\bbm\kt,\CHI\kt,{w\kt})\nabla{\bbm\kt}
\big){\cdot}\mathbf n=q_{\rm s,\tau}^k,
\end{align}
\end{subequations}
starting from $k=1$ by using  
\begin{align}\label{IC2}
\!\!{\bfu}_\tau^0=
{\bfu}_0,\quad\ {\bfu}_\tau^{-1}=
{\bfu}_0{-}\tau\mathbf v_0,\quad\ \bbm_\tau^0=\bbm_0,
\quad\ \CHI_\tau^0=\CHI_0,\quad\ {w}_\tau^0=\omega(m_0,\vartheta_0).
\end{align}
An
important feature of the scheme \eqref{TGMd} is that it decouples to 
{three} boundary-value problems, which (after a further spatial 
discretisation) can advantageously be used in a numerical treatment and 
which is advantageously used even to show existence of
approximate solutions:

\begin{lemma}[{\sc Existence of the discrete solution}]\label{lem-1}
Let \eqref{DQ} 
hold and $\tau>0$ be small enough, {cf.\ \eqref{tau-small} below.}
Then, for any $k=1,...,T/\tau$, \eqref{TGMd} possesses a solution
$\bfu\kt\in 
H^1(\Omega;\R^3)$, $\bbm\kt\in H^1(\Omega;\R^N)$, $\xi\kt\in L^2(\Omega;\R^N)$,
$\CHI\kt,\,\mu\kt,\,w\kt\in H^1(\Omega)$
such that {$\CHI\kt\ge0$} and $w\kt\ge0$.
\end{lemma}

\noindent{\it Proof}.
The first boundary-value problem arising by the decoupling is (\ref{TGMd}a,b)
with (\ref{BC-t}a,b), and it leads to the minimization of the functional:
\begin{align}\nonumber
&(\bfu,\bbc)\mapsto\int_\Omega
{\delta_K(\bbc)}+\varphi_{01}(\bfeps(\bfu),\bbc,\CHI\kkt)
+\frac\lambda2|\nabla\bbc|^2
+\frac{\tau^2}2\varrho\Big|\frac{\bfu-2\bfu\kkt+\bfu_\tau^{k-2}}{\tau^2}\Big|^2
\\&\qquad\quad\nonumber
+\frac\tau2\mathbb D\bfeps\big(\frac{\bfu{-}\bfu\kkt}\tau\big){:}
\bfeps\big(\frac{\bfu{-}\bfu\kkt}\tau\big)
+\alpha\frac\tau2\Big|\frac{\bbc{-}\bbc\kkt}\tau\Big|^2
+\tau\zeta\Big(\frac{\bbc{-}\bbc\kkt}\tau\Big)
\\
&\qquad\quad
+{\bfsigma_{\rm a}}(\bbm\kkt,w\kkt){:}\bfeps(\bfu)
\label{minimization-u-m}
+\bbs_{\rm a}(\bbm\kkt,w\kkt){\cdot}\bbc
+\bff\kt{\cdot}\bfu\,\d x+\int_\Gamma\bff_{\rm s,\tau}^k{\cdot}\bfu\,\d S.
\end{align}
Due to \eqref{def-of-phi2} and \eqref{semiconvexity-phi0}, 
$\varphi_{12}(\cdot,\cdot,\CHI\kkt)$ 
and thus the whole functional \eqref{minimization-u-m} 
are strictly convex, for $\tau>0$ 
sufficiently small, {namely for 
\begin{align}\label{tau-small}
\tau\le\tau_1:=
\begin{cases}\displaystyle{\min\Big(T,\frac{\alpha^2}
{|\inf\partial_{\bbc\bbc}^2\varphi_1|^2}\Big)}
&\text{if }\inf\partial_{\bbc\bbc}^2\varphi_1<0,\\[-.3em]
T&\text{otherwise}.\end{cases}
\end{align}}
Therefore, there exists a unique minimizer 
$(\bfu\kt,\bbc\kt)\!\in\!H^1(\Omega;\R^3)\!\times\!H^1(\Omega;\R^N)$.
{By standard arguments, cf.\ e.g.\ \cite[Chap.\,2 and 4]{Roubicek2013},}
this minimizer {when completed by $\xi\kt\!\in\!N_K(\bbc\kt)$}
is a weak solution to (\ref{TGMd}a,b)--(\ref{BC-t}a,b). {Moreover,
$\xi\kt\!\in\!L^2(\Omega;\R^N)$ can be shown by the same arguments
as in the Lemma~\ref{lem-1+} below.}

Then one can solve \eqref{gilbmod-disc}--\eqref{bc34+}, which represents a 
semi-linear diffusion equation. We observe that we can eliminate $\mu\kt$ 
because obviously
\begin{align}\label{mu-m-chi}
\nabla\mu\kt=\partial_{\CHI\bbm}^2{\varphi_1}(\bbm\kt,\CHI\kt)\nabla\bbm\kt
+\partial_{\CHI\CHI}^2{\varphi_1}(\bbm\kt,\CHI\kt)\nabla\CHI\kt,
\end{align} 
which leads us to abbreviate
\begin{subequations}\label{def-of-Ms}
\begin{align}
&{\mathsf M}_1(\bfeps,\bbm,\CHI,w):=
\partial_{\CHI\CHI}^2{\varphi_1}(\bbm,\CHI)\mathsf M(\bfeps,\bbm,\CHI,w),
\\&\mathsf M_2(\bfeps,\bbm,\CHI,w):=
\mathsf M(\bfeps,\bbm,\CHI,w)\otimes\partial_{\CHI\bbm}^2{\varphi_1}(\bbm,\CHI).
\end{align}
\end{subequations}
Then \eqref{gilbmod-disc}--\eqref{bc34+} transforms to the semi-linear boundary-value problem:
\begin{subequations}\label{BVP-for-chi}
\begin{align}
&
\deltax-\textrm{div}
\big({\mathsf M}_1(\bfeps(\bfu\kt),\bbm\kt,\CHI\kt,w\kkt)\nabla\CHI\kt
\!+{\mathsf M}_2(\bfeps(\bfu\kt),\bbm\kt,\CHI\kt,w\kkt)\nabla\bbm\kt\big)=0
\intertext{on $\Omega$ together with the boundary condition on $\Gamma$:}
&
\big(
{\mathsf M}_1(\bfeps(\bfu\kt),
\bbm\kt,\CHI\kt,w\kkt)\nabla\CHI\kt+{\mathsf M}_{2}(\bfeps(\bfu\kt),
\bbm\kt,\CHI\kt,w\kkt)\nabla\bbm\kt\big){\cdot}\mathbf n
=\bff_{\rm s,\tau}^k.
\end{align}
\end{subequations}
Due to the fully implicit discretisation of 
$\partial_{\CHI}{\varphi_1}(\bbm,\CHI)$ which is needed for the a-priori 
estimates in Lemma~\ref{lem-2} below, via \eqref{def-of-Ms} we inevitably 
obtain the dependence of $\mathsf M_{1,2}(\bfeps(\bfu\kt),\bbm\kt,\CHI\kt,w\kkt)$ on $\CHI\kt$ so that the problem \eqref{BVP-for-chi} unfortunately does not have any potential. 
Anyhow, thanks to \eqref{pos1}, the 2nd-order tensor $\mathsf M_1$ is uniformly positive definite. 
Also, by \eqref{bbg} and \eqref{Mbound}, 
$\mathsf M_2$ is bounded. As a consequence, the nonlinear operator 
$A:H^1(\Omega)\mapsto H^1(\Omega)^*$ defined by
\begin{align}\nonumber
\big\langle A(\CHI),\bbv\big\rangle:= \int_\Omega\frac1\tau\CHI\bbv&+
\Big(\mathsf M_{1}(\bfeps(\bfu\kt),\bbm\kt,\CHI,w\kkt)\nabla\CHI
\\[-.3em]&\ \ +
\mathsf  M_{2}(\bfeps(\bfu\kt),\bbm\kt,\CHI,w\kkt)\nabla\bbm\kt\Big)
{\cdot}\nabla\bbv\,\d x
\end{align}
is coercive. Thanks to Assumption \eqref{growth1}, we have that $A$ 
is also weakly continuous, existence of a solution
$\CHI\kt\in H^1(\Omega)$ can be obtained using the Galerkin method
and Brouwer fixed-point theorem; of course, the obtained solution needs not be unique. {Testing \eqref{BVP-for-chi} by $[\CHI\kt]^-$ and
using \eqref{bbg} implies $\CHI\kt\ge0$ provided $\CHI\kkt\ge0$, which
yields non-negativity of the hydrogen concentration recursively for
any $k=1,...,T/\tau$ by using \eqref{IC-ass-2}.}

Let us also note that from \eqref{mu-m-chi} we obtain also 
$\nabla\mu\kt\in L^2(\Omega;\R^3)$. In particular, we have simply 
{both $\mathbbm D\bfeps(\deltau){:}\bfeps(\deltau)/(1{+}\tau|\bfeps(\deltau)|^2)\in L^\infty(\Omega)$ and}
$\mathsf M(\bfeps(\bfu\kt),\bbm\kt,\CHI\kt,w\kkt)
\nabla\mu\kt{\cdot}\nabla\mu\kt/(1{+}\tau|\nabla\mu\kt|^2)\in L^\infty(\Omega)$,
and thus the right-hand side of \eqref{heatequation-disc} is in $L^2(\Omega)$.
Therefore, eventually, we are to solve \eqref{heatequation-disc}--\eqref{BC-t-2-2}, 
which represents a semilinear heat-transfer equation with the right-hand side 
in $H^1(\Omega)^*$. The only nonlinearity is due to the $w$-dependence of 
$\mathsf{L}(\bfeps(\bfu\kt),\bbm\kt,\CHI\kt,w)$, $\bfsigma_{\rm a}(\bbm\kkt,w)$, 
and $\bbs_{\rm a}(\bbm\kkt,w)$. The later two are needed to guarantee 
$w\kt\ge0$. Anyhow, since this nonlinearity is of lower order, we can pass 
through it by compactness and strong convergence. Thus, it suffices for us to 
check coercivity of the underlying operator. To this aim, we test 
\eqref{heatequation-disc} by $w\kt$. The terms on the right-hand side containing 
$w\kt$ are estimated standardly by using H\"older's and Young's inequalities, 
and using the qualifications {(\ref{DQ}l,m).}
Having coercivity, we see that there exists at least one solution. 
Moreover, this solution satisfies $w\kt\ge0$, which can be seen by testing 
$\eqref{heatequation-disc}$ by the negative part of  $w\kt$ and using that ${\bfsigma_{\rm a}}(\bbm,w)=0$ and ${{\bbs_{\rm a}}}(\bbm,w)=0$ for $w\le0$. 
Note that to have such non-negativity it is important to have $w\kt$ in the $\mathbbm K$-term on the left-hand side, as well as in the nonlinear terms $\sigma_{\rm a}$ and $\bbs_{\rm a}$ on the right-hand
side. 
\QED

\medskip

Let us define the piecewise affine interpolant $u_\tau$  by 
\begin{subequations}
\begin{align}
&\bfu_\tau(t):=
\frac{t-(k{-}1)\tau}\tau\bfu\kt
+\frac{k\tau-t}\tau \bfu\kkt
\quad\text{ for $t\in[(k{-}1)\tau,k\tau]\ $}
\intertext{with $\ k=0,...,T/\tau$. Besides, we define also the 
backward piecewise constant interpolant $\bar u_\tau$ and $\underline u_\tau$
by}
&\bar\bfu_\tau(t):=\bfu\kt,
\qquad\qquad\qquad\qquad\text{ for $t\in((k{-}1)\tau,k\tau]\ $,\ \ 
$k=1,...,T/\tau$},
\\
&\underline\bfu_\tau(t):=\bfu\kkt,
\qquad\qquad\qquad\quad\ \text{for $t\in[(k{-}1)\tau,k\tau)\ $,\ \ 
$k=1,...,T/\tau$}.
\intertext{Similarly, we define also $m_\tau$, $\bar m_\tau$, 
$\underline m_\tau$, $\bar\vartheta_\tau$, $\vartheta_\tau$,
$\bar g_\tau$, $\bar\bff_{\rm b,\tau}$, 
etc. We will also need the piecewise affine interpolant of the 
(piecewise constant) velocity $\frac{\partial\bfu_\tau}{\partial t}$,
which we denote by $\big[\frac{\partial\bfu_\tau}{\partial t}\big]^{\rm i}$, i.e.}
&\big[\DT\bfu_\tau\big]^{\rm i}(t):=
\frac{t{-}(k{-}1)\tau}\tau\,\frac{\bfu\kt{-}\bfu\kkt}\tau
+\frac{k\tau{-}t}\tau
\,\frac{\bfu\kkt{-}\bfu_\tau^{k-2}}\tau
\ \text{ for $t\in((k{-}1)\tau,k\tau]$}.
\end{align}
\end{subequations}
Note that $\DDT\bfu_\tau^{\rm i}:=\frac{\partial}{\partial t}
\big[\DT\bfu_\tau\big]^{\rm i}$ is piecewise constant with the values 
$\frac{\bfu_\tau^k{-}2\bfu_\tau^{k-1}{+}\bfu_\tau^{k-2}}{\tau^2}$ on the 
particular subintervals $((k{-}1)\tau,k\tau)$.

In terms of these interpolants, we can write the approximate system 
\eqref{TGMd} in a more ``condensed'' form closer to the desired continuous 
system \eqref{BVP-t}, namely:
\begin{subequations}\label{BVP-disc}
\begin{align}
&\r\DDT\bfu_\tau^{\rm i}
-{\rm div}\big(
{\bbC(\bfeps(\bar\bfu_\tau){-}\bfepstr\bar\bbc_\tau)}
+{\bfsigma_{\rm a}}(\underline{\bbm}_\tau,\underline{w}_\tau)
\label{TGMd-1-comp}
+\mathbb D\bfeps(\DT\bfu_\tau)\big)={\bar\bff_\tau},
\\
&
\nonumber
%{}
\alpha\DT\bbc_\tau{+}\partial\zeta(\DT\bbc_\tau)
{-}\lambda\Delta\bar\bbc_\tau
{+}\partial_\bbc{\varphi_1}(\bar\bbc_\tau,\underline\CHI_\tau)
{+}\bfepstr^\top\bbC(\bfepstr\bar\bbc_\tau{-}\bfeps(\bar\bfu_\tau))
\\
\label{TGMd-2-comp}
&\hspace{6em}
%{}
+{\bbs_{\rm a}}({\underline{\bbm}_\tau},\underline{w}_\tau){+}\bar\xi_\tau\ni0
\hspace{3.1em}\text{ with }\ \ \ \ \bar\xi_\tau\in N_K^{}(\bar\bbc_\tau),
\\&\label{gilbmod-disc2}
\DT\CHI_\tau-\textrm{div}\big({\mathsf M}(\bfeps(\bar\bfu_\tau),
\bbm\kt,\bar\CHI_\tau,\underline{w}_\tau)\nabla\bar\mu_\tau\big)=0\ \ \ 
\ \text{ with }\ \ \ \
\bar\mu_\tau=\partial_{\CHI}{\varphi_1}(\bar\bbm_\tau,\bar\CHI_\tau),
\\
&
\DT w_\tau-\mathrm{div}\big(\mathsf{K}(\bfeps(\bar\bfu_\tau),
\bar\bbm_\tau,\bar\CHI_\tau,\underline{w}_\tau)\nabla \bar w_\tau
+\mathsf{L}(\bfeps(\bar\bfu_\tau),\bar\bbm_\tau,\bar\CHI_\tau,\underline{w}_\tau)
\nabla\bar\bbm_\tau\big)\nonumber
\\[-.3em]&\hspace{6em}=\big(\bfsigma_{\rm a}(\underline\bbm_\tau,\bar w_\tau)+
\frac{\mathbbm D\,\bfeps(\DT\bfu_\tau)
\big){:}\bfeps(\DT\bfu_\tau)}{{1+\tau|\bfeps(\DT\bfu_\tau)|^2}}
+\big(\bbs_{\rm a}({\underline{\bbm}_\tau},\bar w_\tau)
+\alpha\DT\bbm_\tau\big){\cdot}\DT\bbm_\tau\nonumber
\\\label{heatequation2}&\hspace{6em}
+\zeta(\DT\bbm_\tau)
+
\frac{\mathsf M(\bfeps(\bar\bfu_\tau),\bar\bbm_\tau,\bar\CHI_\tau,\underline{w}_\tau)
\nabla\bar\mu_\tau{\cdot}\nabla\bar\mu_\tau}{1+\tau|\nabla\bar\mu_\tau|^2}
+{\bar q_\tau},
\end{align}
\end{subequations}
for $k=1,...,T/\tau$, together with the boundary conditions
\begin{subequations}\label{BC-disc}\begin{align}
&\big(\bbC(\bfeps(\bar\bfu_\tau){-}\bfepstr\bar\bbc_\tau)
+{\bfsigma_{\rm a}}(
\underline{\bbm}_\tau,
\underline{w}_\tau)
+\mathbb D\bfeps(\DT\bfu_\tau)
\big)\mathbf n={\bar\bff_{\rm s,\tau}},
\label{BC-t-1}
\\
&
\frac{\partial\bar\bbc_\tau}{\partial\mathbf n}=0,\qquad\
\label{bc342}
{\mathsf M}(\bfeps(\bar\bfu_\tau),\bar\bbm\kt,
\underline\CHI_\tau,\underline{w}_\tau)\nabla\bar\mu_\tau
\cdot\mathbf n={\bar h_{\rm s,\tau}},
\\
&\label{BC-t-2-3}
\big(
\mathsf{K}(\bfeps(\bar\bfu_\tau),\bar\bbm_\tau,\bar\CHI_\tau,\underline{w}_\tau)
\nabla \bar w_\tau+\mathsf{L}(\bfeps(\bar\bfu_\tau),\bar\bbm_\tau,\bar\CHI_\tau,\underline{w}_\tau)
\nabla\bar\bbm_\tau\big){\cdot}\mathbf n={\bar q_{\rm s,\tau}}.
\end{align}\end{subequations}

\begin{lemma}[{\sc First estimates}]\label{lem-2}
\slshape
Let again the assumptions of Lemma~\ref{lem-1} hold. Then, for some $C$ and ${C_r}$ independent of $\tau>0$, 
\begin{subequations}\label{apriori-I}
\begin{align}\label{apriori-Ia}
&\big\|\bfu_\tau\big\|_{W^{1,\infty}(I;L^2(\O;\R^3))\,\cap\,
H^1(I;H^1(\O;\R^3))}\le C,
\\\label{apriori-Ia+}
&\big\|\bbm_\tau\big\|_{L^\infty(I;H^1(\O;\R^N))\,\cap\,
H^1(I;L^2(\O;\R^N))\,\cap\,L^\infty(Q;\R^N)}\le C,
\\\label{apriori-Ia++}
&\big\|\bar\CHI_\tau\big\|_{L^\infty(I;H^1(\O))}\le C,
\\\label{apriori-Ia+++}
&\big\|\bar\mu_\tau\big\|_{L^\infty(I;H^1(\O))}\le C,
\\\label{apriori-Ib}
&\big\|\bar w_\tau\big\|_{L^\infty(I;L^1(\O))}\le C,
\\\label{est-of-nabla-w}
&\big\|\nabla\bar w_\tau\big\|_{{L^r(Q;\R^3)}}\le C_r
\qquad\text{ for any }\ 1\le r<5/4.
\end{align}
\end{subequations}
\end{lemma}

\noindent{\it Proof}.
{The strategy is to test the particular equations in \eqref{TGMd} respectively
by $\deltau$,  $\deltam$, $\mu\kt$, and $\frac12$.
For (\ref{TGMd}a,b),} we note that 
a standard convexity argument yields:
\begin{align}
  \label{eq:2}
&\varrho[\dtt\kt]^2\bfu{\cdot}\dtt\kt\bfu+
\bbC(\bfeps(\bfu\kt){{-}\bfepstr\bbc\kt)}
){:}\dtt\kt\bfeps
{+\big(\bfepstr^\top\bbC(\bfepstr\bbc\kt{-}\bfeps(\bfu\kt))
+\xi\kt\big)
{\cdot}\dtt\kt\bbc
+\lambda\nabla\bbc\kt{:}\nabla\dtt\kt\bbc
}
\nonumber\\&\,
\ge\frac\varrho2|\dtt\kt\bfu|^2+\frac12\bbC
(\bfeps(\bfu\kt){-}\bfepstr\bbc\kt){:}(\bfeps(\bfu\kt){-}\bfepstr\bbc\kt)
+\frac\lambda2|\nabla\bbc\kt|^2
+\delta_K(\bbc\kt)
\nonumber\\&\,
-\frac\varrho2|\dtt\kkt\bfu|^2\!-\frac12\bbC
(\bfeps(\bfu\kkt){-}\bfepstr\bbc\kkt){:}(\bfeps(\bfu\kkt){-}\bfepstr\bbc\kkt)
-\frac\lambda2|\nabla\bbc\kkt|^2\!-\delta_K(\bbc\kkt).
\end{align}
Owing the our {equi-semiconvexity assumption 
\eqref{semiconvexity-phi0} on $\varphi_1(\cdot,\CHI)$, we also have:
\begin{align}\label{eq:4}
  &\partial_\bbc\varphi_1(\bbc\kt,\CHI\kkt){\cdot}\dtt\kt\bbc
=\Big(\partial_\bbc\varphi_1(\bbc\kt,\CHI\kkt)
+\alpha\frac{\bbc\kt}{\sqrt\tau}\Big){\cdot}\dtt\kt\bbc
-\alpha\frac {\bbc\kt}{\sqrt\tau}{\cdot}\dtt\kt\bbc
\nonumber\\
&\qquad\ge
\frac1\tau\Big(\varphi_1(\bbc\kt,\CHI\kkt)+\alpha\frac{(\bbc\kt)^2}{2\sqrt\tau}-
\varphi_1(\bbc\kkt,\CHI\kkt)-\alpha\frac{(\bbc\kkt)^2}{2\sqrt\tau}
\Big)
-\alpha\sqrt\tau \frac {\bbc\kt}\tau{\cdot}\dtt\kt\bbc
\nonumber\\
&\qquad=
\frac{\varphi_1(\bbc\kt,\CHI\kkt)-\varphi_1(\bbc\kkt,\CHI\kkt)}\tau
-\alpha\frac{\sqrt \tau}2\left|\dtt\kt\bbc
\right|^2
\end{align}
provided $0<\tau\le\tau_1$ with $\tau_1$ from \eqref{tau-small}.
}
Now, 
we add 
{the tested equations} together. We then use \eqref{eq:2}--\eqref{eq:4}, 
and still the convexity \eqref{phi0-coerc}
of $\CHI\mapsto\varphi_1(\bbm,\CHI)$ to deduce the estimate
\begin{align}\nonumber
&\int_\Omega\frac12w\kt+\frac\r2\big|\dtt\kt\bfu\big|^2
+\varphi_{12}(\bfeps\kt,\bbm\kt,\CHI\kt)
+\frac\lambda2|\nabla\bbm\kt|^2
\,\d x
\\&\nonumber
\ \ +\tau\sum_{j=1}^k\int_\Omega\mathbb D\dtt_\tau^j\bfeps:\dtt_\tau^j\bfeps
+
\frac{2{-}\sqrt\tau}2\alpha\big|\dtt_\tau^j\bbm\big|^2\Big)
+\frac12{\mathsf M}(\bfeps_\tau^j,
\bbm_\tau^j,\CHI_\tau^{j-1},w_\tau^{j-1})
\nabla\mu_\tau^j{\cdot}\nabla\mu_\tau^j
\d x\d t
 \\\nonumber&\ \le\int_\Omega\frac12w_0+\frac\r2|\bfv_0|^2
 +\varphi_{12}(\bfeps_0,\bbm_0,\CHI_0)
 +\frac\lambda2|\nabla\bbm_0|^2
 %+{\frac\tau\eta|{\bfeps}(\bfu_{0,\tau})|^\eta}
\,\d x
\\\nonumber&
\ \ +\tau\sum_{j=1}^k\int_\Omega
 \bff^j_\tau{\cdot}\dtt_\tau^j\bfu\,+\frac12q_\tau^j
+\Big(\frac12\bfsigma_{\rm a}(\bbm_\tau^{j-1},w_\tau^j)
 -\bfsigma_{\rm a}(\bbm_\tau^{j-1},w_\tau^{j-1})
 \Big)
 \cdot\dtt_\tau^j\bfeps
 \\
&
\ \
+\Big(\frac12\bbs_{\rm a}({\bbm_\tau^{j-1}}, w_\tau^{j})-
 \bbs_{\rm a}({\bbm_\tau^{j-1}}, w_\tau^{j-1})
\Big){\cdot}\dtt_\tau^j\bbc\,\d x
\label{eq:7}
+\int_\Gamma\bff^j_{\rm s,\tau}{\cdot}\dtt_\tau^j\bfu+
  h^j_{\rm s,\tau}{\cdot}\dtt_\tau^j\mu_\tau+\frac12 q^j_{\rm s,\tau}\d S
\end{align}
{where we used the abbreviation $\varphi_{12}$ from \eqref{def-of-phi12}
with $\varphi_2$ from \eqref{def-of-phi2}.}

We remark that our semi-implicit scheme has benefited from the cancellation
of the terms $\pm\frac1\tau\varphi_1(\bbc\kt,\CHI\kkt)$ 
under this test by time-differences, which is a more general phenomenon 
to be understood as the fractional-step method, here combined with the 
semiconvexity, cf.\ \cite[Remarks~8.24-8.25]{Roubicek2013}.
{Now, by \eqref{q1}, using  H\"older inequality, 
and recalling that $w_\tau^k\ge 0$, we obtain the estimate:\\[-.9em]
\begin{align}\label{gf}
&\int_\Omega 
 \Big(\frac12\bfsigma_{\rm a}(\bbm_\tau^{j-1},w_\tau^j)
 -\bfsigma_{\rm a}(\bbm_\tau^{j-1},w_\tau^{j-1})
 \Big){:}\dtt_\tau^j\bfeps\,\d x\nonumber\\[-.5em]
&\qquad\le C_\epsilon+C_\epsilon\int_\Omega\varphi_1(\bbm_\tau^{j-1},\CHI_\tau^{j-1})+w_\tau^j+ w_\tau^{j-1}+|\bfeps_\tau^j|^2\, \d x+\epsilon \int_\Omega |\dtt_\tau^j\bfeps|^2\, \d x,
\end{align}
where $\epsilon$ is an arbitrarily small number, and $C_\epsilon$ depends on 
$\epsilon$.
A similar estimate holds also for the terms multiplying $\dtt_\tau^j\bbc$ 
on the right-hand side of \eqref{eq:7}. 
We  now can adsorb the discrete time derivatives into the left-hand side, and 
use a 
discrete Gronwall inequality. 
Thanks to \eqref{pos1} and \eqref{phi0-coerc}, we have 
$\varphi_1(\bbm,\CHI)\ge\epsilon\CHI^2$ 
and the a-priori bound $\bbm\kt\in K$.
This gives the estimates (\ref{apriori-I}a-c,e). The 
estimate \eqref{apriori-Ia+++} follows from the relation 
(cf.\ also \eqref{mu-m-chi}):
\begin{align}\label{est-of-nabla-concentration}
\nabla\bar\CHI_\tau=
\big[\partial_{\CHI\CHI}
^2{\varphi_1}(\bar\bbm_\tau,\bar\CHI_\tau)\big]^{-1}
\big(\nabla\bar\mu_\tau-\partial_{\CHI\bbm}^2{\varphi_1}(\bar\bbm_\tau,\bar\CHI_\tau)
\nabla\bar\bbm_\tau\big).
\end{align} 
Eventually, we observe that the right-hand sides of \eqref{heatequation2} and 
\eqref{BC-t-2-3} are bounded in $L^1(Q)$ and $L^1(\Sigma)$, respectively. For 
this, we need, in particular, assumptions \eqref{q1} and \eqref{q2}. Then one 
can use the $L^1$-theory for the heat equation  to obtain the remaining 
estimate \eqref{est-of-nabla-w}, see \cite{Boccardo1997}{}. {}To this goal, {like in \cite{FePeRo09ESPT},}
one is to perform the test of \eqref{heatequation-disc}
by $\varpi'(w\kt)$, where {}$\varpi(w\kt):=((1{+}{w\kt})^{1-\e}\!-1)/(\e{-}1)$ with $\e>0$, and sum it 
for $k=1,...,T/\tau$. Comparing to standard technique, see for instance \cite{Podio-Guidugli2010,Roubicek2011,Roubivcek2010}, the only non-standard estimates is due to the $\mathsf{L}$-term, as in \cite{Roubivcek2012}, which requires assumption \eqref{q3}, cf.\ also \cite[Sect.12.9]{Roubicek2013}.
}
\QED

\begin{lemma}[{\sc Further estimates}]\label{lem-1+}
\slshape
Under the assumption of Lemma~\ref{lem-1}, for 
some constant $C$ independent of $\tau$, it also holds:
\begin{subequations}\label{apriori-II+}
\begin{align}\label{apriori-IIc}
&\big\|\r\DDT\bfu_\tau^{\rm i}
\big\|_{L^2(I;H^1(\O;\R^3)^*)}\le C,
\\\label{a-priori-IId}
&\big\|\DT{w}_\tau\big\|_{L^1(I;H^3(\O)^*)}\le C,
\\\label{a-priori-IIe}
&\big\|\DT{\CHI}_\tau\big\|_{L^2(I;H^1(\O)^*)}\le C,
\\\label{est-of-Delta-m}
&\big\|\Delta\bar\bbm_\tau\big\|_{L^2(Q;\R^N)}\le C,
\\\label{est-of-xi}
&{\big\|\bar\xi_\tau\big\|_{L^2(Q;\R^N)}\le C.}
\end{align}
\end{subequations}
\end{lemma}

\noindent{\it Proof}.
The ``dual'' estimates (\ref{apriori-II+}a-c) can be obtained routinely as 
a consequence of the previously derived ones by using the equations 
(\ref{BVP-disc}a,d,e) with the corresponding boundary conditions
(\ref{BC-disc}a,d,e). As $\zeta$ is finite, $\partial\zeta$ is bounded, 
and from 
{\eqref{growth-of-phi1} and \eqref{q1}}
and the already proved estimates (\ref{apriori-I}a-c,e),
 we have, for some $C$ finite,
\begin{align*}
\forall\,\bar r_\tau\!\in\!\partial\zeta(\DT\bbc_\tau){+}\alpha\DT\bbc_\tau
{+}\partial_\bbm{\varphi_1}(\bar\bbm_\tau,\underline\CHI_\tau)
{+}\partial_\bbm{\varphi_2}(\bfeps(\bar\bfu_\tau),\bar\bbm_\tau)
{+}{\bbs_{\rm a}}({\underline{\bbm}_\tau},\underline{w}_\tau){:}
\ \ \big\|\bar r_\tau\big\|_{L^2(Q;\R^N)}\!\le C.
\end{align*}
We can now prove (\ref{apriori-II+}d,e), imitating an abstract 
procedure like in \cite{LuScSt08GSPF}. We write 
{\eqref{TGMd-2-comp} as\\[-1.9em] 
\begin{align}\label{Delta-m+}
\lambda\Delta\bar\bbc_\tau-\bar\xi_\tau=\bar r_\tau\ \ \text{ with }\ \ 
\bar\xi_\tau\in N_K^{}(\bar\bbc_\tau).
\end{align}
We test \eqref{Delta-m+} by $\Delta\bar\bbc_\tau$ and use 
the monotonicity of the set-valued mapping $N_K^{}$,
which  ensures (when written very formally) that
\begin{align}\nonumber
&\!\int_\Omega\!\bar\xi_\tau{\cdot}\Delta\bar\bbc_\tau\,\d x=
\int_\Omega\!N_K^{}(\bar\bbc_\tau){\cdot}\Delta\bar\bbc_\tau\,\d x=
\int_\Gamma\! N_K^{}(\bar\bbc_\tau)
\frac{\partial\bar\bbc_\tau}{\partial\mathbf n}\,\d S
-\int_\Omega\!\nabla N_K^{}(\bar\bbc_\tau){:}\nabla\bar\bbc_\tau\,\d x
\\[-.3em]\label{L2-regularity}&\qquad\ \
=-\int_\Omega 
\partial N_K^{}(\bar\bbc_\tau)\nabla\bar\bbc_\tau{:}\nabla\bar\bbc_\tau\,\d x
=-\int_\Omega \partial^2\delta_K(\bar\bbc_\tau)
\nabla\bar\bbc_\tau{:}\nabla\bar\bbc_\tau\,\d x\le0.
\end{align}
Of course, the positive-definiteness of the Jacobian $\partial^2\delta_K$
of the nonsmooth convex function $\delta_K$ is indeed very formal and
the rigorous proof needs a smoothening argument.
In fact, one can use an exterior penalty $\delta_{K,\epsilon}^{}(\bbc):=
\epsilon^{-1}\min_{\widetilde\bbc\in K}|\bbc{-}\widetilde\bbc|^2$
(i.e.\ Yosida's approximation of $\delta_K^{}$) 
and consider the Dirichlet boundary-value problem 
$\lambda\Delta\bar\bbc_{\tau,\epsilon}-\delta_{K,\epsilon}'(\bar\bbc_{\tau,\epsilon})
=\bar r_\tau$ with the boundary condition $\bar\bbc_{\tau,\epsilon}=\bar\bbc_\tau$
on $\Sigma$ which ensures $\delta_{K,\epsilon}'(\bar\bbc_{\tau,\epsilon})=0$ on 
$\Sigma$ so that the boundary term arising by the test by 
$\Delta\bbc_{\tau,\epsilon}$ disappears, likewise already in 
\eqref{L2-regularity}, and the limit with $\epsilon\to0$ is then easy.
Therefore, by this estimate, we obtain 
$\lambda\|\Delta\bar\bbc_\tau\|_{L^2(Q;\R^N)}\le
\|\bar r_\tau\|_{L^2(Q;\R^N)}$ so that \eqref{est-of-Delta-m} is proved,
and thus also the bound \eqref{est-of-xi} for 
$\bar\xi_\tau=\lambda\Delta\bar\bbc_\tau-\bar r_\tau$
is proved.}
\QED

\medskip

\begin{proposition}[{\sc Convergence for $\tau\to0$}]\label{prop-conv}
\slshape
Let again the assumption of Lemma~\ref{lem-1} hold.
Then there is a subsequence such that
\begin{subequations}\label{15.-strong}
\begin{align}
\label{15.-u-strong}
&\bfu_\tau\to\bfu&&\text{strongly in }\ H^1(I;H^1(\O;\R^3)),
\\\label{15.-m}
&\bbm_\tau\to \bbm&&\text{strongly in }\ {H^1(I;L^2(\O;\R^N))},
\\\label{15.-chi}
&\bar\CHI_\tau\to\CHI&&{\text{strongly in }\ L^r(Q)\ \text{ with any }\ 
1\le r<6},
\\
\label{15.-theta-strong}
&\bar{w}_\tau\to {w}\ \ \&\ \ \underline{w}_\tau\to {w}\!\!\!\!\!\!\!\!
&&\text{strongly in }\ L^s(Q)\ \text{ with any }\ 
{1\le s<5/3},
\\\label{15.-mu-strong}
&{\bar\mu_\tau\to\mu}&&\text{strongly in }\ L^2(I;H^1(\O)),
\\\label{15.-xi-weak}
&{\bar\xi_\tau\to\xi}&&\text{weakly in }\ L^2(Q;\R^N),
\end{align}\end{subequations}
and any {$(\bfu,\bbm,\CHI,w,\mu,\xi)$} obtained in this way 
is a weak solution 
to the system \eqref{BVP-t}--\eqref{bcc}
in accord with Definition~\ref{def}
which also preserves the total energy {as well as $\CHI\ge0$ and $w\ge0$}
as claimed in Theorem~\ref{thm}. {Moreover, if $\Omega$ is smooth, then 
also 
\begin{subequations}\label{15.-strong+}
\begin{align}
\label{15.-m+}
&\!\bar\bbm_\tau\to \bbm&&&&\quad\ \text{strongly in }\ {L^2(I;H^1(\O;\R^N))},&&&&
\\\label{15.-chi+}
&\!\bar\CHI_\tau\to \CHI&&&&\quad\ \text{strongly in }\ {L^2(I;H^1(\O))}.
\end{align}\end{subequations}
}
\end{proposition}

\noindent{\it Proof}. 
For lucidity, we divide the proof into {nine} steps.

\medskip

\noindent
\emph{Step 1: Selection of a converging subsequence.} By Banach's selection 
principle, we select a weakly* converging subsequence with respect to the 
norms from the estimates \eqref{apriori-I} and \eqref{apriori-II+}.
{By Aubin-Lions' theorem, 
from \eqref{apriori-Ia++} and \eqref{a-priori-IIe}, one gets \eqref{15.-chi}.}
Moreover,  a (generalized) Aubin-Lions theorem, cf.\ 
\cite[Corollary~7.9]{Roubicek2013}, based on \eqref{est-of-nabla-w} 
and \eqref{a-priori-IId} which makes $\DT{\bar w}_\tau$ bounded as an 
$H^3(\Omega)^*$-valued measure on $[0,T]$,
interpolated further with the estimate \eqref{apriori-Ib},
gives the first strong convergence \eqref{15.-theta-strong}. The second one
follows analogously.

{
Further,
$\CHI\ge0$ and $w\ge0$ 
is 
inherited from 
$\CHI_\tau\ge0$ and $w_\tau\ge0$ proved in Lemma~\ref{lem-1}.
If
 $\Omega$ is smooth, from \eqref{est-of-Delta-m} 
one has $\bar\bbm_\tau$ bounded in $L^2(I;H^2(\Omega;\R^N))$,
so by Aubin-Lions' theorem one gets \eqref{15.-m+} and then,
from \eqref{est-of-nabla-concentration}, one gets also \eqref{15.-chi+}.
}

\medskip

\noindent
\emph{Step 2: Convergence in the semilinear mechanical  part.}
Equation (\ref{balmech}) is obviously semilinear, and therefore
the weak convergence is sufficient to obtain it in the limit from the
corresponding equations (\ref{TGMd-1-comp}).
In particular, $\varrho\DDT\bfu$ is in duality with $\DT\bfu$:
\begin{align}\label{quality-of-u}
\DT\bfu\in L^2(I;H^1(\O;\R^3))\quad \text{and}\quad 
\r\DDT\bfu\in L^2(I;H^1(\O;\R^3)^*).
\end{align}
By 
Rellich's theorem 
we have the continuous and the compact embeddings 
${H^1(I;L^2(\Omega))\cap L^\infty(I;H^1(\Omega))}
\subset H^1(Q)\Subset L^2(Q)$ so that
$\bbm_\tau\to\bbm$ strongly in $L^2(Q;\R^N)$ 
and then also $\underline\bbm_\tau\to\bbm$ strongly in $L^2(Q;\R^N)$ 
because $\|\underline\bbm_\tau{-}\bbm_\tau\|_{L^2(Q;\R^N)}
=3^{-1/2}\tau\|\DT\bbm_\tau\|_{L^2(Q;\R^N)}\to0$, cf.\ 
\cite[Rem.\,8.10]{Roubicek2013}. Together with \eqref{15.-theta-strong},
we can pass through the nonlinearities
$\bfsigma_{\rm a}$
and $\bbs_{\rm a}$. {Similarly also 
$\bar\bbm_\tau\to\bbm$ strongly in $L^2(Q;\R^N)$, which we will use later.} 

\medskip

\noindent
\emph{Step 3: Strong convergence of $\bfeps(\bfu_\tau)$.}
As the nonlinearities $\partial_\bbm\varphi_2$, 
$\mathsf{M}$, $\mathsf{K}$, and $\mathsf{L}$
may depend on $\bfeps$, we need first to prove strong convergence of 
$\bfeps(\bfu_\tau)$.
Using the equation for the discrete approximation, and the limit equation 
proved in Step 2, 
we can write:
\begin{align}\nonumber
&\int_Q\!\bbC\bfeps(\bar\bfu_\tau{-}\bfu){:}\bfeps(\bar\bfu_\tau{-}\bfu)\,\d x\d t
\\[-.3em]&\qquad\qquad
\le 
\int_Q\!\varrho\DDT\bfu_\tau^{\rm i}
{\cdot}(\bfu{-}\bar\bfu_\tau)
+\bfsigma_{\rm a}(\underline\bbm_\tau,\underline\CHI_\tau){:}
\bfeps(\bfu{-}\bar\bfu_\tau)
+\bbC\bfeps(\bfu){:}\bfeps(\bfu{-}\bar\bfu_\tau)\,\d x\d t
\label{strong-e(u)}
\end{align}
where we used the monotonicity of $u\mapsto\mathbb D\bfeps(\DT u)$
if the initial condition is fixed; 
cf.\ \cite{Roubicek2013} for details about the time discretisation. 
The goal is to pass the right-hand side of \eqref{strong-e(u)} to 0. 
We use
Aubin-Lions' Theorem to obtain strong convergence of $\DT\bfu_\tau$ in 
$L^2(Q;\R^3)$, and by the Rellich compactness theorem  
also strong convergence of $\bar\bfu_\tau(T)$ in 
$L^2(\Omega;\R^3)$, which allows us to pass to the limit:
\begin{align*}
&\lim_{\tau\to 0}
\int_Q\!\varrho\DDT\bfu_\tau^{\rm i}
{\cdot}(\bfu{-}\bar\bfu_\tau)\,\d x\d t=\lim_{\tau\to 0}
\bigg(\int_\O\varrho \DT\bfu_\tau(0)\cdot\bar\bfu_\tau(\tau)
-\varrho\DT\bfu_\tau(T)\cdot\bar\bfu_\tau(T)\,\d x
\\[-.3em]&\hspace{15em}
+\int_\tau^T\!\!\!\int_\Omega\varrho\DT\bfu_\tau(\cdot-\tau){\cdot}
\DT\bfu_\tau\,\d x\d t
+\int_Q\varrho\DDT\bfu_\tau^{\rm i}{\cdot}\bfu\,\d x\d t\bigg)
\\&\hspace{7em}=\int_\O\varrho \DT\bfu(0){\cdot}\bfu(0)
-\varrho\DT\bfu(T){\cdot}\bfu(T)\,\d x
+\int_Q\varrho|\DT\bfu_\tau|^2+\varrho\DDT\bfu{\cdot}\bfu\,\d x\d t=0;
\end{align*}
the former equation is by discrete by-part summation, cf.\ e.g.\ 
\cite[Remark~11.38]{Roubicek2013}, while the later equality simply 
follows by reversing integration by parts;
here \eqref{quality-of-u} is used. By \eqref{apriori-Ia++} 
and \eqref{a-priori-IIe}, we have 
$\underline\CHI_\tau\to\CHI$ weakly* in $L^\infty(I;H^1(\Omega))$,
and $\underline{\DT\CHI}_\tau$ bounded as an $H^1(\Omega)^*$-valued
measure on $[0,T]$, which gives $\underline\CHI_\tau\to\CHI$ strongly in 
$L^2(Q)$ by a (generalized) Aubin-Lions theorem, cf.\ 
\cite[Corollary~7.9]{Roubicek2013}. 
Then the other terms in \eqref{strong-e(u)}
clearly converge 0. 
Altogether, we proved that 
the right-hand side of \eqref{strong-e(u)} converges to 0, 
which eventually shows 
$\bar\bfeps_\tau\to\bfeps$ in $L^2(Q;\R^{3\times3})$.

\medskip

\emph{Step 4: Limit passage in the micromechanical 
inequality.} 
{F}rom \eqref{TGMd-2-comp}, we have
\begin{subequations}\begin{align}\nonumber
&\forall \bbv\!\in\!L^2(I;H^1(\Omega;\R^N)){:}\,\ 
\int_Q\!
\big(\alpha\DT\bbm_\tau
+\bbs_{\rm a}(\underline\bbm_\tau,\underline\CHI_\tau)
+
\partial_\bbc\varphi_{12}(\bar\bfeps_\tau,\bar\bbc_\tau,\underline\CHI_\tau)
{+\bar\xi_\tau}\big){\cdot}(\bbv{-}\DT\bbc_\tau)
\\[-.5em]\label{mm-ineq-1}
&\hspace{11em}
+\lambda\nabla\bar\bbm_\tau{:}(\nabla\bbv{-}\nabla\DT\bbm_\tau)
+\zeta(\bbv)\,\d x\d t\ge\!\int_Q\!
\zeta(\DT\bbm_\tau)\,\d x\d t
\\[-.2em]&\label{mm-ineq-2}
{\forall\bbv\!\in\!L^2(Q;\R^N),\ \ \bbv\!\in\!K\text{ a.e.}:\quad 
\int_Q\!\bar\xi_\tau{\cdot}(\bbv{-}\bar\bbc_\tau)\,\d x\d t\ge0.}
\end{align}\end{subequations}
{The limit passage in \eqref{mm-ineq-2} is easy because 
$\bar\xi_\tau\to\xi$ weakly in $L^2(Q;\R^N)$ and 
$\bar\bbc_\tau\to\bbc$ strongly in $L^2(Q;\R^N)$ has already been proved
in Step 2; thus $\xi\in N_K^{}(\bbc)$ is shown. Now we can make a limit passage in 
\eqref{mm-ineq-1}.} 
Here on the left-hand side we have collected all terms that need to be handled
through a continuity or a weak upper semicontinuity arguments,
while the right-hand side is to be treated by weak {}lower {}semicontinuity.
We benefit from the strong convergence of $\underline\bbm_\tau$ (and
similarly also of $\bar\bbm_\tau$) 
shown in Step~2 and of $\underline\CHI_\tau$ and 
$\bar\bfeps_\tau$ proved in Step 3. The only nontrivial limit 
passage which, however, leads us directly to \eqref{def-of-m} is:
\begin{align}\nonumber
\limsup_{\tau\to0}\int_Q\!-\lambda\nabla\bar\bbm_\tau{\cdot}\nabla\DT\bbm_\tau
\,\d x\d t&\le\int_\Omega\!\frac\lambda2\big|\nabla\bbm_0\big|^2\,\d x
-\liminf_{\tau\to0}\int_\Omega\!\frac\lambda2\big|\nabla\bbm(T)\big|^2\,\d x
\\&\le\int_\Omega\frac\lambda2\big|\nabla\bbm_0\big|^2
-\frac\lambda2\big|\nabla\bbm(T)\big|^2\,\d x.
\end{align}
{Eventually, the limit in $\int_Q\bar\xi_\tau{\cdot}\DT\bbc_\tau\,\d x\d t$
is simple because, for any $\bar\xi_\tau\in N_K^{}(\DT\bbc_\tau)$,
this integral equals $\int_\Omega\delta_K(\xi_\tau(T))-\delta_K(\xi_0)\,\d x=0
=\int_\Omega\delta_K(\xi(T))-\delta_K(\xi_0)\,\d x
=\int_Q\xi{\cdot}\DT\bbc\,\d x\d t$ since we already know $\xi\in N_K^{}(\bbc)$;
note that \eqref{IC-ass-1} has been used here. Thus
\eqref{def-of-m} is proved.
}

\medskip

\noindent
\emph{Step 5: Limit passage in the diffusion equation.}
Again, all strong convergences we already have proved in Steps 2 and 3
are to be used. Then the limit passage in the semi-linear 
equation \eqref{gilbmod-disc} is simple. 

\medskip

\noindent
\emph{Step 6: Mechanical/chemical energy preservation.} We 
balance the kinetic and stored energy integrated over the domain:\\[-1.5em]
\[
\mathcal E(t):=\int_\O\frac\varrho2\big|\DT\bfu(t)\big|^2
+{\varphi_{12}(\bfeps(\bfu(t),\bbm(t),\CHI(t))}
+\frac\lambda2\big|\nabla\bbm(t)\big|^2\,\d x.
\]
{as we actually did in \eqref{balance3} with $\nu=0$}, i.e.\\[-1.5em]
\begin{align}\nonumber
&\mathcal  E(T)-\mathcal E(0)=\int_Q\!\bff{\cdot}\DT\bfu+q\,\d x\d t
-\int_Q\!\big(\mathbb D\bfeps(\DT\bfu){+}
\bfsigma_{\rm a}(\bbm,w)\big){:}\bfeps(\DT\bfu)+
\big(\alpha\DT\bbm{+}\bbs_{\rm a}(\bbm,w)\big){\cdot}\DT\bbm
\\[-.3em]\label{engr-equality}&\qquad\qquad\quad\ \ +\zeta(\DT\bbm)
+\mathsf{M}(\bfeps(\bfu),\bbm,\CHI,w)\nabla\mu{\cdot}\nabla\mu\,\d x\d t
+\int_\Sigma\!\bff_{\rm s}{\cdot}\DT\bfu{+}q_{\rm s}{+}\mu h_{\rm s}\,\d S\d t.
\end{align}
This is standardly achieved by testing the  mechanical-chemical equations 
(\ref{BVP-t}a,c), and the inclusion (\ref{BVP-t}b), respectively by 
$\DT\bfu$, $\mu$, and $\DT\bbm$, and by using the chain rule to integrate with 
respect to $t$. In order for these tests to be legal, we need $\varrho\DDT\bfu$
to be in duality with $\DT\bfu$, which has already been proved, 
cf.\ \eqref{quality-of-u}.
In particular, we make use of\\[-1.5em]
\[
\int_0^T\!\big\langle\varrho\DDT\bfu,\DT\bfu\big\rangle\,\d t=
\int_\Omega \frac\varrho2\big|\DT\bfu(T)\big|^2
-\frac\varrho2\big|\DT\bfu(0)\big|^2\,{\rm d}x.
\]
{Further, we need $\Delta\bbm\in L^2(Q;\R^N)$ to have \eqref{by-part-for-m}
at our disposal; for this, the assumptions
\eqref{growth-of-phi1} and \eqref{q1} together with the estimates 
(\ref{apriori-I}a-c,e) are used.}
{Also, $\DT\CHI\in L^2(I;H^1(\Omega)^*)$ is in duality with 
$\mu\in L^2(I;H^1(\Omega))$
as well as $\DT\bbm\in L^2(Q;\R^N)$ is in duality with 
$\partial_\bbm\varphi_1(\bbm,\CHI)\in L^2(Q;\R^N)$, 
cf.\ \eqref{a-priori-IIe} with \eqref{15.-mu-strong}
and \eqref{apriori-Ia+} with the assumption \eqref{growth-of-phi1}
with $\bbm\in L^\infty(Q;\R^N)$ and $\CHI\in L^\infty(I;L^6(\Omega))$, so that 
we can rigorously execute the formula \eqref{fundamental-test}
integrated over $I$, which gives\\[-1.5em]
\begin{equation*}
\int_0^T\bigg(\langle\DT\CHI,\mu\rangle+
\int_\Omega\partial_\bbm\varphi_1(\bbm,\CHI)\DT\bbm\,\d x\bigg)\d t
=\int_\Omega\varphi_1(
\bbm(T),\CHI(T))-\varphi_1(\bbm_0,\CHI_0)\,\d x.
\end{equation*}
Also $\xi\in L^2(Q;\R^N)$ is in duality with $\DT\chi\in L^2(Q;\R^N)$ so that 
$\int_Q\xi\DT\bbm\,\d x\d t$ has a sense and simply equals to 0 because 
$\xi\in\partial\delta_K(\bbm)$ has been proved in Step~4 and because 
$\delta_K(\bbm_0)=0$ is assumed, cf.\ \eqref{IC-ass-1}.
}

\medskip

\noindent
\emph{Step 7: Strong convergence of ${\bfeps}(\DT\bfu_\tau)$, 
$\DT\bbm_\tau$, and 
$\nabla\bar\mu_\tau$.}
Using 
the discrete mechanic-chemical energy imbalance 
{(which is like \eqref{eq:7} except that the 
1/2 of the heat equation \eqref{heatequation2} is not counted),}
and eventually the energy equality \eqref{engr-equality}, 
we can write\goodbreak
\begin{align}\nonumber
&\int_Q \zeta(\DT\bbm)+
\bbD{\bfeps}(\DT\bfu){:}{\bfeps}(\DT\bfu)
+\alpha|\DT\bbm|^2
+\bbM(\bfeps(\bfu),\bbm,\CHI,w)\nabla\mu{\cdot}\nabla\mu\,\d x\d t
\\[-.4em]&\ \nonumber\le
\liminf_{\tau\to0}
\int_Q\!\zeta(\DT\bbm_\tau)+
\bbD{\bfeps}(\DT\bfu_\tau){:}{\bfeps}(\DT\bfu_\tau)
+\alpha|\DT\bbm_\tau|^2
+
\bbM(\bfeps(\bar\bfu_\tau),\bar\bbm_\tau,\bar\CHI_\tau,\underline w_\tau)
\nabla\bar\mu_\tau
{\cdot}\nabla\bar\mu_\tau
\\&\ \nonumber\le\limsup_{\tau\to0}
\int_Q\!\zeta(\DT\bbm_\tau)+
\bbD{\bfeps}(\DT\bfu_\tau){:}{\bfeps}(\DT\bfu_\tau)
+{\big(1{-}\sqrt\tau/2\big)}\alpha|\DT\bbm_\tau|^2
\\[-.5em]&\hspace{15.7em}\nonumber
+\bbM(\bfeps(\bar\bfu_\tau),\bar\bbm_\tau,\bar\CHI_\tau,\underline w_\tau)
\nabla\bar\mu_\tau{\cdot}\nabla\bar\mu_\tau\,\d x\d t
\\\nonumber
&\ \le\limsup_{\tau\to0}\bigg(
\mathcal E(0)
-{\!\int_\O\!\frac\varrho2\big|\DT\bfu_\tau(T)\big|^2\!
+\varphi_{12}(\bfeps(\bfu_\tau(T)),\bbm_\tau(T),\CHI_\tau(T))
+\frac\lambda2\big|\nabla\bbm_\tau(T)\big|^2\d x}
\\[-.4em]\nonumber
&\hspace{4.5em}
-\int_\Sigma\!\bar \bff_{{\rm s},\tau}{\cdot}\DT\bfu_\tau\dS\d t
+\int_Q\!\bar\bff_\tau{\cdot}\DT\bfu_\tau
-\bfsigma_{\rm a}(\underline\bbm_\tau,\underline w_\tau)
{:}\bfeps(\DT\bfu_\tau)
-\bbs_{\rm a}(\underline\bbm_\tau,\underline w_\tau){\cdot}\DT\bbm_\tau
\dx\d t\bigg)
\\[-.3em]\nonumber
&\ \le
\mathcal E(0)-\mathcal E(T)-\int_\Sigma \bff_{\rm s}{\cdot}\DT\bfu\dS\d t
+\int_Q\bff{\cdot}\DT\bfu
-\bfsigma_{\rm a}(\bbm,w){:}\bfeps(\DT\bfu)
-\bbs_{\rm a}(\bbm,w){\cdot}\DT\bbm\,\dx\d t
\\[-.5em]&\ =\int_Q \zeta(\DT\bbm)+\bbD{\bfeps}(\DT\bfu){:}\bfeps(\DT\bfu)
+\alpha|\DT\bbm|^2
+\bbM(\bfeps(\bfu),\bbm,\CHI,w)\nabla\mu{\cdot}\nabla\mu\,\d x\d t.
\label{lim-inf-sup}
\end{align}
Thus we can write ``lim'' and ``='' everywhere in \eqref{lim-inf-sup}
and, together with the already proved weak convergence, 
we obtain the desired strong convergence of $\nabla{\bfeps}(\DT\bfu_\tau)$
and $\DT\bbm_\tau$ and $\nabla\bar\mu_\tau$ in $L^2(Q)$-spaces.
For technical details about the term $\bbM\nabla\mu{\cdot}\nabla\mu$ 
with the nonconstant coefficient 
$\bbM=\bbM(\bfeps(\bfu),\bbm,\CHI,w)$, we refer to 
\cite[Formula (4.25)]{Roubivcek2010}.

\smallskip

\noindent
\emph{Step 8: Limit passage in the heat equation \eqref{heatequation2}.}
Having proved the strong convergence in {Steps 3 and 7}, the right-hand 
side of \eqref{heatequation2} converges strongly in $L^1(Q)$ and 
this limit passage towards the weak solution to
\eqref{heatequation}--\eqref{BC-t-2} is then easy.

\smallskip

\noindent
\emph{Step 9: Total energy preservation, {i.e.\ \eqref{balance3} with 
$\nu=1$}.} 
{We have $\DT{w}\in L^1(I;H^3(\Omega)^*)$, cf.\ \eqref{a-priori-IId}
and realize the already proved identity \eqref{heatequation}, 
which is in duality with the constant 1, we can perform rigorously this
test and sum it with mechanical/chemical energy balance obtained already in 
Step 6.}
\QED

\section{Generalization of the model for electrolytes and fuel cell modeling}\label{sec-fuel-cells}

In a very basic scenario, the above presented model allows for
a relatively simple generalization for a multicomponent, charged 
(i.e.\ ionized), chemically reacting medium undergoing electro-diffusion  
in elastic medium. 
When having in mind \emph{hydrogen fuel cells}, the 
specific situation involves elastic polymeric porous layer with 
negatively-charged dopands and undergoing mechanical deformation/stresses 
e.g.\ due to {\it swelling} through which water H$_2$O, hydrogen ions H$^+$ 
(i.e.\ protons), and {hydronium} ions H$_3$O$^+$
(or, in general, ions of the type H$_{2n+1}$O$_n^+$ also with $n\ge0$)
move by drift and diffusion; we speak about
a polymeric electrolyte membrane (=PEM). This membrane is surrounded by two 
thin layers where catalyzed chemical reactions
take place and another electron-conductive layers called an
anode and a cathode. The mentioned reactions are H$_2\to$2H$^++2$e$^-$ 
(on the layer between the anode and membrane) and 
O$_2$+4H$^+$+4e$^-\to$2H$_2$O (on the layer between the cathode and membrane).
There is vast amount of literature about such fuel cells and their
modeling, cf.\ e.g.\ \cite{Fuhr13MNMF,Kuli10AMFC,ProWet09PEMF} for a survey. 
Similar scenario applies for methanol or ethanol fuel cells, except a different 
chemistry on the anode. All the models seem however to be focused on 
electro-chemistry without taking (thermo)mechanical interactions properly 
into account, sometimes being rather one-dimensional and mostly 
not accompanied by any mathematical analysis. Of course, 
the following outlined model can serve only as an ansatz to
which a lot of concrete data is to be supplied.

The generalization of the above presented model to $m$ diffusive constituents
consists in taking the concentration $\boldsymbol\CHI$ and 
the (now electro-)chemical potential $\boldsymbol\mu$ vector valued.
Moreover, we consider a vector of electric charges  $\mathbf{z}$ of the 
$m$ constituents; some components of $\mathbf{z}$ can be zero.
Further, we consider the vector of chemical-reactions rate 
${\mathbf r}={\mathbf r}(x,\boldsymbol\CHI)$,
electrostatic potential $\phi$ of the self-induced
electric field, an electric 
permittivity $\epsilon=\epsilon(x)$, and $d=d(x)$ concentration of dopands.
The $x$-dependence allows for distinction of particular layers
composing the mentioned fuel cells. The above outlined chemistry 
was only an example - some alternative similar application might be 
e.g.\ methanol fuel cells, cf.\ \cite{DFGJ03PMDM} for a simplified model.

The mass balance \eqref{EEd-t2} together with \eqref{def-of-mu}
augments to  
\begin{subequations}\label{Rosbroeck}
\begin{align}\label{mass-conserv}
&\DT{\boldsymbol\CHI}-
\mathrm{div}\big(\bbM(\bfeps(\bfu),\bbm,\boldsymbol\CHI,\vartheta)
\nabla\boldsymbol\mu\big)={\mathbf r}(\boldsymbol\CHI),
\\
&\boldsymbol\mu=
\partial_\CHI\varphi_1(\bbm,\boldsymbol\CHI)+\mathbf{z}\phi,
\end{align}
where $\phi$ is to solve the (rest of Maxwell system for) 
electrostatics balancing the electrical induction $\epsilon\nabla\phi$ 
as\\[-1.2em]
\begin{align}\label{poisson}
&-\mathrm{div}\big(\epsilon\nabla\phi\big)
=\mathbf{z}{\cdot}\boldsymbol\CHI+d\qquad\qquad
\end{align}
\end{subequations}
with some (here unspecified) boundary conditions. In \eqref{mass-conserv},
$\bbM$ is now a 4-th order symmetric tensor and thus
$[\mathrm{div}(\bbM\nabla\boldsymbol\mu)]_i^{}:=
\sum_{k=1}^3\frac{\partial}{\partial x_k}
\sum_{l=1}^3\sum_{j=1}^m\bbM_{ijkl}
\frac{\partial\boldsymbol\mu_j}{\partial x_l}$.

The essence of this electro-statical augmentation is that the 
test of \eqref{mass-conserv} by $\boldsymbol\mu$ now relies on the 
modification of \eqref{fundamental-test} if integrated over $\Omega$ as 
follows:\\[-1.5em]
\begin{align}\nonumber
\int_\Omega\DT{\boldsymbol\CHI}{\cdot}\boldsymbol\mu\,\d x&=
\int_\Omega\DT{\boldsymbol\CHI}{\cdot}\big(\partial_\CHI\varphi_1(\bbm,\boldsymbol\CHI)+\mathbf{z}\phi\big)\,\d x
\\\nonumber&=
\frac{\d}{\d t}\int_\Omega\varphi_1(\bbm,\boldsymbol\CHI)\,\d x
-\int_\Omega\partial_\bbm\varphi_1(\bbm,\boldsymbol\CHI){\cdot}
\DT{\boldsymbol\mu}-\phi\mathbf{z}{\cdot}\DT{\boldsymbol\CHI}\,\d x
\\\nonumber&=
\frac{\d}{\d t}\int_\Omega\varphi_1(\bbm,\boldsymbol\CHI)\,\d x
-\int_\Omega\partial_\bbm\varphi_1(\bbm,\boldsymbol\CHI){\cdot}
\DT{\boldsymbol\mu}-\phi\,\mathrm{div}\big(\epsilon\nabla\DT\phi\big)\,\d x
\\&=
\frac{\d}{\d t}\int_\Omega\varphi_1(\bbm,\boldsymbol\CHI)
+\frac\epsilon2|\nabla\phi|^2\,\d x
-\int_\Omega\partial_\bbm\varphi_1(\bbm,\boldsymbol\CHI){\cdot}
\DT{\boldsymbol\mu}\,\d x
\end{align}
together with (for simplicity unspecified) term arising from boundary 
conditions for \eqref{poisson}. The energy balance \eqref{balance3}
now involves also the energy of the electrostatic field 
$\frac12\int_\Omega\epsilon|\nabla\phi|^2\,\d x$. The term 
${\mathbf r}(\boldsymbol\CHI){\cdot}\boldsymbol\mu
={\mathbf r}(\boldsymbol\CHI){\cdot}
\partial_\CHI\varphi_1(\bbm,\boldsymbol\CHI)
+{\mathbf r}(\boldsymbol\CHI){\cdot}\mathbf{z}\phi$ 
arising by the mentioned test of \eqref{mass-conserv} is to be
treated by Gronwall's inequality under some growth qualification on
the chemical-reaction rates ${\mathbf r}$. The convergence analysis
imitates \cite{Roub07IINF} as far as the 
electrostatic part concerns. 
Standard modeling of electro-chemical devices with 
sizes substantially larger than the so-called Debye length like fuel cells 
however simplifies the model by considering local electroneutrality, arising
as an asymptotics for $\epsilon\to0$ in \eqref{poisson}, cf.\ e.g.\ 
\cite{Fuhr13MNMF}.

Final note is that, considering $m=2$ and forgetting $\bfeps$, $\bbm$, and 
$\vartheta$, the system \eqref{Rosbroeck} itself represents the
classical Roosbroeck's \emph{drift-diffusion model for semiconductors} 
\cite{Roos50TFEH},
the components of $\boldsymbol\CHI$ being then interpreted as concentrations
of electrons and holes; cf.\ e.g.\ \cite[Sect.\,12.4]{Roubicek2013}.
The generalization presented in this section can thus be also interpreted
in its very special case $m=2$ as a model for thermodynamics of \emph{elastic 
semiconductors}, especially if the mobility tensor $\bbM$ would
be allowed to depend also on the intensity of the electric field $\nabla\phi$.\\

\bibliographystyle{abbrv}
\bibliography{tr-gt-hydro}

\bigskip

\end{sloppypar}
\end{document}